\documentclass[twocolumn]{aastex61}
\shorttitle{Subaru High-$z$ Exploration of Low-Luminosity Quasars (SHELLQs) V}
\shortauthors{Matsuoka et al.}

\begin{document}

\title{Subaru High-{\scriptsize $z$} Exploration of Low-Luminosity Quasars (SHELLQ{\scriptsize s}). V. Quasar Luminosity Function and Contribution to Cosmic Reionization at {\scriptsize $z$}\ = 6}

\correspondingauthor{Yoshiki Matsuoka}
\email{yk.matsuoka@cosmos.ehime-u.ac.jp}

\author{Yoshiki Matsuoka}
\affil{Research Center for Space and Cosmic Evolution, Ehime University, Matsuyama, Ehime 790-8577, Japan.}

\author{Michael A. Strauss}
\affil{Princeton University Observatory, Peyton Hall, Princeton, NJ 08544, USA.}

\author{Nobunari Kashikawa}
\affil{Department of Astronomy, School of Science, The University of Tokyo, Tokyo 113-0033, Japan.}
\affil{National Astronomical Observatory of Japan, Mitaka, Tokyo 181-8588, Japan.}
\affil{Department of Astronomical Science, Graduate University for Advanced Studies (SOKENDAI), Mitaka, Tokyo 181-8588, Japan.}

\author{Masafusa Onoue}
\affil{Max Planck Institut f\"{u}r Astronomie, K\"{o}nigstuhl 17, D-69117, Heidelberg, Germany}
\affil{National Astronomical Observatory of Japan, Mitaka, Tokyo 181-8588, Japan.}
\affil{Department of Astronomical Science, Graduate University for Advanced Studies (SOKENDAI), Mitaka, Tokyo 181-8588, Japan.}

\author{Kazushi Iwasawa}
\affil{ICREA and Institut de Ci{\`e}ncies del Cosmos, Universitat de Barcelona, IEEC-UB, Mart{\'i} i Franqu{\`e}s, 1, 08028 Barcelona, Spain.}

\author{Ji-Jia Tang}
\affil{Institute of Astronomy and Astrophysics, Academia Sinica, Taipei, 10617, Taiwan.}

\author{Chien-Hsiu Lee}
\affil{National Optical Astronomy Observatory, 950 North Cherry Avenue, Tucson, AZ 85719, USA.}
\affil{Subaru Telescope, Hilo, HI 96720, USA.}

\author{Masatoshi Imanishi}
\affil{National Astronomical Observatory of Japan, Mitaka, Tokyo 181-8588, Japan.}
\affil{Department of Astronomical Science, Graduate University for Advanced Studies (SOKENDAI), Mitaka, Tokyo 181-8588, Japan.}

\author{Tohru Nagao}
\affil{Research Center for Space and Cosmic Evolution, Ehime University, Matsuyama, Ehime 790-8577, Japan.}

\author{Masayuki Akiyama}
\affil{Astronomical Institute, Tohoku University, Aoba, Sendai, 980-8578, Japan.}

\author{Naoko Asami}
\affil{Seisa University, Hakone-machi, Kanagawa, 250-0631, Japan.}

\author{James Bosch}
\affil{Princeton University Observatory, Peyton Hall, Princeton, NJ 08544, USA.}

\author{Hisanori Furusawa}
\affil{National Astronomical Observatory of Japan, Mitaka, Tokyo 181-8588, Japan.}

\author{Tomotsugu Goto}
\affil{Institute of Astronomy and Department of Physics, National Tsing Hua University, Hsinchu 30013, Taiwan.}

\author{James E. Gunn}
\affil{Princeton University Observatory, Peyton Hall, Princeton, NJ 08544, USA.}

\author{Yuichi Harikane}
\affil{Institute for Cosmic Ray Research, The University of Tokyo, Kashiwa, Chiba 277-8582, Japan}
\affil{Department of Physics, Graduate School of Science, The University of Tokyo, Bunkyo, Tokyo 113-0033, Japan}

\author{Hiroyuki Ikeda}
\affil{National Astronomical Observatory of Japan, Mitaka, Tokyo 181-8588, Japan.}

\author{Takuma Izumi}
\affil{National Astronomical Observatory of Japan, Mitaka, Tokyo 181-8588, Japan.}

\author{Toshihiro Kawaguchi}
\affil{Department of Economics, Management and Information Science, Onomichi City University, Onomichi, Hiroshima 722-8506, Japan.}

\author{Nanako Kato}
\affil{Graduate School of Science and Engineering, Ehime University, Matsuyama, Ehime 790-8577, Japan.}

\author{Satoshi Kikuta}
\affil{National Astronomical Observatory of Japan, Mitaka, Tokyo 181-8588, Japan.}
\affil{Department of Astronomical Science, Graduate University for Advanced Studies (SOKENDAI), Mitaka, Tokyo 181-8588, Japan.}

\author{Kotaro Kohno}
\affil{Institute of Astronomy, The University of Tokyo, Mitaka, Tokyo 181-0015, Japan.}
\affil{Research Center for the Early Universe, University of Tokyo, Tokyo 113-0033, Japan.}

\author{Yutaka Komiyama}
\affil{National Astronomical Observatory of Japan, Mitaka, Tokyo 181-8588, Japan.}
\affil{Department of Astronomical Science, Graduate University for Advanced Studies (SOKENDAI), Mitaka, Tokyo 181-8588, Japan.}

\author{Robert H. Lupton}
\affil{Princeton University Observatory, Peyton Hall, Princeton, NJ 08544, USA.}

\author{Takeo Minezaki}
\affil{Institute of Astronomy, The University of Tokyo, Mitaka, Tokyo 181-0015, Japan.}

\author{Satoshi Miyazaki}
\affil{National Astronomical Observatory of Japan, Mitaka, Tokyo 181-8588, Japan.}
\affil{Department of Astronomical Science, Graduate University for Advanced Studies (SOKENDAI), Mitaka, Tokyo 181-8588, Japan.}


\author{Hitoshi Murayama}
\affil{Kavli Institute for the Physics and Mathematics of the Universe, WPI, The University of Tokyo,Kashiwa, Chiba 277-8583, Japan.}

\author{Mana Niida}
\affil{Graduate School of Science and Engineering, Ehime University, Matsuyama, Ehime 790-8577, Japan.}

\author{Atsushi J. Nishizawa}
\affil{Institute for Advanced Research, Nagoya University, Furo-cho, Chikusa-ku, Nagoya 464-8602, Japan.}

\author{Akatoki Noboriguchi}
\affil{Graduate School of Science and Engineering, Ehime University, Matsuyama, Ehime 790-8577, Japan.}

\author{Masamune Oguri}
\affil{Department of Physics, Graduate School of Science, The University of Tokyo, Bunkyo, Tokyo 113-0033, Japan}
\affil{Kavli Institute for the Physics and Mathematics of the Universe, WPI, The University of Tokyo,Kashiwa, Chiba 277-8583, Japan.}
\affil{Research Center for the Early Universe, University of Tokyo, Tokyo 113-0033, Japan.}

\author{Yoshiaki Ono}
\affil{Institute for Cosmic Ray Research, The University of Tokyo, Kashiwa, Chiba 277-8582, Japan}

\author{Masami Ouchi}
\affil{Institute for Cosmic Ray Research, The University of Tokyo, Kashiwa, Chiba 277-8582, Japan}
\affil{Kavli Institute for the Physics and Mathematics of the Universe, WPI, The University of Tokyo,Kashiwa, Chiba 277-8583, Japan.}

\author{Paul A. Price}
\affil{Princeton University Observatory, Peyton Hall, Princeton, NJ 08544, USA.}

\author{Hiroaki Sameshima}
\affil{Koyama Astronomical Observatory, Kyoto-Sangyo University, Kita, Kyoto, 603-8555, Japan.}

\author{Andreas Schulze}
\affil{National Astronomical Observatory of Japan, Mitaka, Tokyo 181-8588, Japan.}

\author{Hikari Shirakata}
\affil{Department of Cosmosciences, Graduates School of Science, Hokkaido University, N10 W8, Kitaku, Sapporo 060-0810, Japan.}

\author{John D. Silverman}
\affil{Kavli Institute for the Physics and Mathematics of the Universe, WPI, The University of Tokyo,Kashiwa, Chiba 277-8583, Japan.}

\author{Naoshi Sugiyama}
\affil{Kavli Institute for the Physics and Mathematics of the Universe, WPI, The University of Tokyo,Kashiwa, Chiba 277-8583, Japan.}
\affil{Graduate School of Science, Nagoya University, Furo-cho, Chikusa-ku, Nagoya 464-8602, Japan.}

\author{Philip J. Tait}
\affil{Subaru Telescope, Hilo, HI 96720, USA.}

\author{Masahiro Takada}
\affil{Kavli Institute for the Physics and Mathematics of the Universe, WPI, The University of Tokyo,Kashiwa, Chiba 277-8583, Japan.}

\author{Tadafumi Takata}
\affil{National Astronomical Observatory of Japan, Mitaka, Tokyo 181-8588, Japan.}
\affil{Department of Astronomical Science, Graduate University for Advanced Studies (SOKENDAI), Mitaka, Tokyo 181-8588, Japan.}

\author{Masayuki Tanaka}
\affil{National Astronomical Observatory of Japan, Mitaka, Tokyo 181-8588, Japan.}
\affil{Department of Astronomical Science, Graduate University for Advanced Studies (SOKENDAI), Mitaka, Tokyo 181-8588, Japan.}

\author{Yoshiki Toba}
\affil{Department of Astronomy, Kyoto University, Sakyo-ku, Kyoto, Kyoto 606-8502, Japan.}
\affil{Institute of Astronomy and Astrophysics, Academia Sinica, Taipei, 10617, Taiwan.}

\author{Yousuke Utsumi}
\affil{Kavli Institute for Particle Astrophysics and Cosmology, Stanford University, CA 94025, USA.}

\author{Shiang-Yu Wang}
\affil{Institute of Astronomy and Astrophysics, Academia Sinica, Taipei, 10617, Taiwan.}

\author{Takuji Yamashita}
\affil{Research Center for Space and Cosmic Evolution, Ehime University, Matsuyama, Ehime 790-8577, Japan.}



\begin{abstract}
We present new measurements of the quasar luminosity function (LF) at $z \sim 6$, over an unprecedentedly wide range of the rest-frame ultraviolet luminosity 
$M_{1450}$ from $-30$ to $-22$ mag.
This is the fifth in a series of publications from the Subaru High-$z$ Exploration of Low-Luminosity Quasars (SHELLQs) project, which 
exploits the deep multi-band imaging data produced by the Hyper Suprime-Cam (HSC) Subaru Strategic Program survey.
The LF was calculated with a complete sample of 110 quasars at $5.7 \le z \le 6.5$, which includes 48 SHELLQs quasars discovered over 650 deg$^2$, and 63 brighter quasars 
discovered by the Sloan Digital Sky Survey and the Canada-France-Hawaii Quasar Survey (including one overlapping object).
This is the largest sample of $z \sim 6$ quasars with a well-defined selection function constructed to date, and has allowed us to detect significant flattening 
of the LF at its faint end.
A double power-law function fit to the sample yields a faint-end slope $\alpha = -1.23^{+0.44}_{-0.34}$, a bright-end slope $\beta = -2.73^{+0.23}_{-0.31}$, a break magnitude 
$M_{1450}^* = -24.90^{+0.75}_{-0.90}$, and a characteristic space density $\Phi^* = 10.9^{+10.0}_{-6.8}$ Gpc$^{-3}$ mag$^{-1}$.
Integrating this best-fit model over the range $-18 < M_{1450} < -30$ mag, 
quasars emit ionizing photons at the rate of $\dot{n}_{\rm ion} = 10^{48.8 \pm 0.1}$ s$^{-1}$ Mpc$^{-3}$ at $z = 6.0$. 
This is less than 10 \% of the critical rate necessary to keep the intergalactic medium ionized, which indicates that quasars are not a major contributor to
cosmic reionization.
\end{abstract}

\keywords{dark ages, reionization, first stars --- galaxies: active --- galaxies: high-redshift --- intergalactic medium --- quasars: general --- quasars: supermassive black holes}



\section{Introduction} \label{sec:intro}

The first billion years of the Universe, corresponding to redshift $z > 5.7$, have been the subject of major observational and theoretical studies in the last few decades.
The first generation of stars, galaxies, and supermassive black holes (SMBHs) are thought to have formed during this epoch, and the Universe 
became reionized during that time, most likely due to the ionizing photons from these light sources.
A large number of high-$z$ galaxies and galaxy candidates
have been identified up to $z \sim 10$ and beyond, and the evolution of the galaxy luminosity function (LF) has been intensively studied 
\citep[e.g.,][]{bouwens11, bouwens15, mcleod16, oesch16, oesch18, ishigaki18}.
\citet{robertson15} demonstrated that these high-$z$ galaxies produced sufficient quantities of ionizing photons to dominate the reionization process, based on the {\it Planck} measurements of the cosmic microwave background polarization \citep{planck16} and an assumed value of the Lyman continuum escape fraction.

The search for high-$z$ quasars\footnote{Throughout this paper, ``high-$z$" denotes $z > 5.7$, where the cosmic age is less than a billion years and objects are observed 
as $i$-band dropouts in the Sloan Digital Sky Survey (SDSS) filter system \citep{fukugita96}.} has also undergone significant progress in the recent years, thanks to the advent of wide-field (1,000-deg$^2$ class) multi-band red-sensitive imaging surveys 
such as the SDSS \citep{york00}, the Canada-France-Hawaii Telescope Legacy Survey (CFHTLS), the Panoramic Survey Telescope \& Rapid Response System 1 \citep[Pan-STARRS1;][]{chambers16},
and the United Kingdom Infrared Telescope (UKIRT) Infrared Deep Sky Survey 
\citep[UKIDSS;][]{lawrence07}.
At the time of writing of this paper, there are 242, 145, 18, and 2 quasars reported in the literature at redshifts beyond $z$ = 5.7, 6.0, 6.5, and 7.0, respectively.
The two highest-$z$ quasars were found at $z = 7.09$ \citep{mortlock11} and $z = 7.54$ \citep{banados18}.
The quasar LF at $z = 6$ has been measured with the complete samples of quasars 
from the SDSS \citep{jiang16} and the Canada-France-Hawaii Quasar Survey \citep[CFHQS;][]{willott10} based on the CFHTLS.
However, the above measurements were limited mostly to $M_{1450} < -24$ mag where the LF is approximated by a single power-law, with only a single CFHQS quasar known
at a fainter magnitude ($M_{1450} = -22.2$ mag).
Thus it has remained unclear whether or not the LF has a break, and what the faint-end slope is if the break exists.
This is a critical issue, since the faint-end shape of the LF reflects a more typical mode of SMBH growth than probed by luminous quasars, and it has a direct impact on
the estimate of the quasar contribution to cosmic reionization.

In the past few years, there have been several attempts to find low-luminosity quasars at $z \sim 6$.
\citet{kashikawa15} found two quasars (one of which may in fact be a galaxy) with $M_{1450} \sim -23$ mag over 6.5 deg$^2$ imaged by Suprime-Cam \citep{miyazaki02},
a former-generation wide-field camera on the Subaru 8.2-m telescope.
The number densities derived from these two (or one) quasars and the faintest CFHQS quasar may point to flattening of the faint-end LF, 
but the small sample size hampered accurate measurements of the LF shape.
\citet{onoue17} took over the analysis of the above Suprime-Cam data,
but found no additional quasars, confirming the number density measured by \citet{kashikawa15}.
On the other hand, \citet{giallongo15} reported {\it Chandra} X-ray detection of five very faint active galactic nuclei (AGNs) at $z \sim 6$, 
with $-19 \le M_{1450} \le -21$ mag,
over 170 arcmin$^2$ of the Great Observatories Origins Deep Survey \citep[GOODS;][]{giavalisco04} field.
This surprisingly high detection rate could indicate 
a significant AGN contribution to cosmic reionization.
However, their results have been challenged by a number of independent deep X-ray studies, finding much lower number densities of faint AGNs \citep[e.g.,][]{weigel15, cappelluti16, vito16, ricci17,parsa18}.
A high number density of high-$z$ faint AGNs may also be in tension with the epoch of \ion{He}{2} reionization inferred from observations \citep{daloisio17, khaire17, mitra18}.

There have also been extensive efforts to measure the quasar LF at lower redshifts, e.g., at $z \sim 4$ \citep[][]{glikman11, ikeda11, masters12} and at $z \sim 5$ \citep[][]{ikeda12, mcgreer13, yang16}.
Recently \citet{kulkarni18} re-analyzed a large sample of quasars compiled from the above and other papers, and reported very bright break magnitudes ($M_{1450}^* < -27$ mag) 
with steep faint-end slopes at $4 \le z \le 6$.
On the other hand, more recent data reaching $\sim$1 mag fainter than the previous measurements seem to suggest that the LF breaks at fainter magnitudes both at $z \sim 4$ \citep{akiyama18}
and $z \sim 5$ \citep[][see the discussion in \S \ref{sec:LF} of this paper]{mcgreer18}.

This paper presents new measurements of the quasar LF at $z \sim 6$, exploiting a complete sample of 110 quasars at $5.7 \le z \le 6.5$.
The sample includes 48 low-luminosity quasars recently discovered by 
the Subaru High-$z$ Exploration of Low-Luminosity Quasars \citep[SHELLQs;][]{paperI} project.
SHELLQs rests on the Subaru Strategic Program (SSP) survey \citep{aihara18_survey} with Hyper Suprime-Cam \citep[HSC;][]{miyazaki18}, a wide-field 
camera mounted on the Subaru telescope.
We are carrying out follow-up spectroscopy of high-$z$ quasar candidates imaged by the HSC, and have so far identified 150 candidates 
over 650 deg$^2$, which include 74 high-$z$ quasars, 25 high-$z$ luminous galaxies, 6 [\ion{O}{3}] emitters at $z \sim 0.8$, and 45 Galactic cool dwarfs 
\citep[][Matsuoka et al. 2018c, in preparation]{paperI, paperII, paperIV}.
We are also carrying out near-infrared (IR) spectroscopy and Atacama Large Millimeter/ submillimeter Array (ALMA) observations of the discovered objects.
The first ALMA results were published in \citet{izumi18}, and further results are in preparation.

This paper is organized as follows.
In \S \ref{sec:sample} we describe our quasar sample to establish the LF, drawn from the SDSS, the CFHQS, and the SHELLQs.
The completeness of the SHELLQs quasar selection is evaluated in \S \ref{sec:completeness}.
The binned and parametric LFs are presented and discussed in \S \ref{sec:LF}, and the quasar contribution to cosmic reionization is estimated in  \S \ref{sec:reionization}.
A summary appears in \S \ref{sec:summary}.
We adopt the cosmological parameters $H_0$ = 70 km s$^{-1}$ Mpc$^{-1}$, $\Omega_{\rm M}$ = 0.3, and $\Omega_{\rm \Lambda}$ = 0.7.
All magnitudes are presented in the AB system \citep{oke83}, and are corrected for Galactic extinction \citep{schlegel98}.
In what follows, we refer to $z$-band magnitudes with the AB subscript (``$z_{\rm AB}$"), while redshift $z$ appears without a subscript.

\section{Quasar Sample} \label{sec:sample}

\begin{figure}
\epsscale{1.2}
\plotone{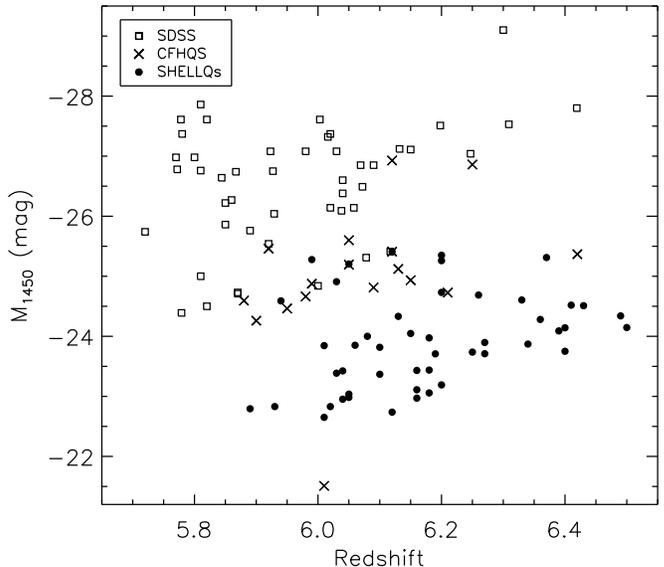}
\caption{The complete quasar sample used in this work, taken from the SDSS (squares), the CFHQS (crosses), and the SHELLQs (dots).
The absolute magnitudes ($M_{1450}$) of the CFHQS quasars have been re-measured in a way consistent with that of the SDSS and SHELLQs (see the text).
\label{fig:sample}}
\end{figure}

We derive the quasar LF with a complete sample of 110 quasars at $5.7\le z \le 6.5$, as summarized in Table \ref{tab:sample} and plotted in Figure \ref{fig:sample}.
These quasars are drawn from the SDSS, the CFHQS, and the SHELLQs, which roughly cover the bright, middle, and faint portions of the magnitude range we probe ($-22 < M_{1450} < -30$ mag),
respectively\footnote{
The present measurements do not include the bright quasars discovered by the Pan-STARRS1 \citep{banados16}, 
whose selection completeness has not been published yet.}.
Table \ref{tab:LFcalc} lists the number of objects in each $M_{1450}$ bin used for the LF calculation, and the corresponding survey volumes ($V_{\rm a}$; see below).

\begin{deluxetable*}{llllllllllll}
\tablecaption{Complete quasar sample \label{tab:sample}}
\tabletypesize{\scriptsize}
\tablehead{
\colhead{Name} &  \colhead{$z$} & \colhead{$M_{1450}$} & \colhead{SC}  &
\colhead{Name} &  \colhead{$z$} & \colhead{$M_{1450}$} & \colhead{SC}  &
\colhead{Name} &  \colhead{$z$} & \colhead{$M_{1450}$} & \colhead{SC} 
} 
\startdata
$J000239.40+255034.8$ & 5.82 & $-27.61$ & 1a & $J092721.82+200123.6$ & 5.77 & $-26.78$ & 1a & $J151248.71+442217.5$ & 6.18 & $-23.06$ & 3  \\
$J000552.33-000655.7$ & 5.85 & $-25.86$ & 1c & $J095740.40+005333.7$ & 6.05 & $-22.98$ & 3  & $J151657.87+422852.9$ & 6.13 & $-24.33$ & 3  \\
$J000825.77-062604.6$ & 5.93 & $-26.04$ & 1b & $J100401.37+023930.9$ & 6.41 & $-24.52$ & 3  & $J152555.79+430324.0$ & 6.27 & $-23.90$ & 3  \\
$J002806.57+045725.3$ & 6.04 & $-26.38$ & 1b & $J103027.09+052455.0$ & 6.31 & $-27.53$ & 1a & $J154552.08+602824.0$ & 5.78 & $-27.37$ & 1a \\
$J003311.40-012524.9$ & 6.13 & $-25.12$ & 2a & $J104433.04-012502.1$ & 5.78 & $-27.61$ & 1a & $J154505.62+423211.6$ & 6.50 & $-24.15$ & 3  \\
$J005006.67+344522.6$ & 6.25 & $-26.86$ & 2a & $J104845.05+463718.4$ & 6.20 & $-27.51$ & 1a & $J160253.98+422824.9$ & 6.09 & $-26.85$ & 1a \\
$J005502.91+014618.3$ & 5.98 & $-24.66$ & 2a & $J105928.61-090620.4$ & 5.92 & $-25.46$ & 2a & $J162331.80+311200.6$ & 6.25 & $-27.04$ & 1a \\
$J010013.02+280225.8$ & 6.30 & $-29.10$ & 1a & $J113717.72+354956.9$ & 6.03 & $-27.08$ & 1a & $J163033.89+401209.7$ & 6.06 & $-26.14$ & 1b \\
$J010250.64-021809.9$ & 5.95 & $-24.46$ & 2a & $J113753.64+004509.7$ & 6.40 & $-24.14$ & 3  & $J164121.64+375520.5$ & 6.05 & $-25.60$ & 2a \\
$J012958.51-003539.7$ & 5.78 & $-24.39$ & 1c & $J114338.34+380828.7$ & 5.81 & $-26.76$ & 1a & $J205321.77+004706.8$ & 5.92 & $-25.54$ & 1c \\
$J013603.17+022605.7$ & 6.21 & $-24.73$ & 2a & $J114648.42+012420.1$ & 6.27 & $-23.71$ & 3  & $J205406.50-000514.4$ & 6.04 & $-26.09$ & 1c \\
$J014837.64+060020.0$ & 5.92 & $-27.08$ & 1a & $J114632.66-015438.2$ & 6.16 & $-23.43$ & 3  & $J210054.62-171522.5$ & 6.09 & $-24.81$ & 2a \\
$J020258.21-025153.6$ & 6.03 & $-23.39$ & 3  & $J114816.64+525150.3$ & 6.42 & $-27.80$ & 1a & $J211951.89-004020.1$ & 5.87 & $-24.73$ & 1c \\
$J020332.38+001229.4$ & 5.72 & $-25.74$ & 1c & $J115221.27+005536.6$ & 6.37 & $-25.31$ & 3  & $J214755.42+010755.5$ & 5.81 & $-25.00$ & 1c \\
$J020611.20-025537.8$ & 6.03 & $-24.91$ & 3  & $J120103.02+013356.4$ & 6.06 & $-23.85$ & 3  & $J220132.07+015529.0$ & 6.16 & $-22.97$ & 3  \\
$J021013.19-045620.8$ & 6.43 & $-24.51$ & 3  & $J120246.37-005701.7$ & 5.93 & $-22.83$ & 3  & $J220417.92+011144.8$ & 5.94 & $-24.59$ & 3  \\
$J021627.81-045534.1$ & 6.01 & $-21.51$ & 2b & $J120737.43+063010.1$ & 6.04 & $-26.60$ & 1b & $J221644.47-001650.1$ & 6.10 & $-23.82$ & 3  \\
$J021721.59-020852.6$ & 6.20 & $-23.19$ & 3  & $J120859.23-020034.8$ & 6.2 & $-24.73$ & 3  & $J221917.22+010249.0$ & 6.16 & $-23.11$ & 3  \\
$J022743.29-060530.3$ & 6.20 & $-25.26$ & 3  & $J121503.42-014858.7$ & 6.05 & $-23.04$ & 3  & $J222309.51+032620.3$ & 6.05 & $-25.20$ & 3  \\
$J023930.24-004505.3$ & 5.82 & $-24.50$ & 1c & $J121721.34+013142.6$ & 6.20 & $-25.35$ & 3  & $J222827.83+012809.5$ & 6.01 & $-22.65$ & 3  \\
$J030331.41-001912.9$ & 6.08 & $-25.31$ & 1c & $J121905.34+005037.5$ & 6.01 & $-23.85$ & 3  & $J222847.71+015240.5$ & 6.08 & $-24.00$ & 3  \\
$J031649.87-134032.3$ & 5.99 & $-24.88$ & 2a & $J124340.81+252923.9$ & 5.85 & $-26.22$ & 1a & $J222901.65+145709.0$ & 6.15 & $-24.93$ & 2a \\
$J035349.73+010404.6$ & 6.07 & $-26.49$ & 1c & $J125051.93+313021.9$ & 6.15 & $-27.11$ & 1a & $J223644.58+003256.9$ & 6.4 & $-23.75$ & 3  \\
$J081054.32+510540.1$ & 5.80 & $-26.98$ & 1a & $J125757.47+634937.2$ & 6.02 & $-26.14$ & 1b & $J223947.47+020747.5$ & 6.26 & $-24.69$ & 3  \\
$J081827.39+172251.8$ & 6.02 & $-27.37$ & 1a & $J130608.25+035626.3$ & 6.02 & $-27.32$ & 1a & $J224237.55+033421.6$ & 5.88 & $-24.59$ & 2a \\
$J083400.88+021146.9$ & 6.15 & $-24.05$ & 3  & $J131911.29+095051.3$ & 6.13 & $-27.12$ & 1b & $J225205.44+022531.9$ & 6.12 & $-22.74$ & 3  \\
$J083525.76+321752.6$ & 5.89 & $-25.76$ & 1b & $J135012.04-002705.2$ & 6.49 & $-24.34$ & 3  & $J225538.04+025126.6$ & 6.34 & $-23.87$ & 3  \\
$J083643.86+005453.2$ & 5.81 & $-27.86$ & 1a & $J140028.80-001151.4$ & 6.04 & $-22.95$ & 3  & $J230422.97+004505.4$ & 6.36 & $-24.28$ & 3  \\
$J084035.09+562419.9$ & 5.84 & $-26.64$ & 1a & $J140319.13+090250.9$ & 5.86 & $-26.27$ & 1b & $J230735.36+003149.3$ & 5.87 & $-24.71$ & 1c \\
$J084119.52+290504.4$ & 5.98 & $-27.08$ & 1b & $J140646.90-014402.5$ & 6.10 & $-23.37$ & 3  & $J231038.88+185519.7$ & 6.00 & $-27.61$ & 1a \\
$J084229.43+121850.5$ & 6.07 & $-26.85$ & 1a & $J140629.13-011611.1$ & 6.33 & $-24.61$ & 3  & $J231546.58-002357.9$ & 6.12 & $-25.41$ & 1c \\
$J084431.60-005254.6$ & 6.25 & $-23.74$ & 3  & $J141111.27+121737.3$ & 5.93 & $-26.75$ & 1a & $J231802.80-024634.0$ & 6.05 & $-25.19$ & 2a \\
$J084408.61-013216.5$ & 6.18 & $-23.97$ & 3  & $J141728.67+011712.4$ & 6.02 & $-22.83$ & 3  & $J232514.25+262847.6$ & 5.77 & $-26.98$ & 1a \\
$J085048.25+324647.9$ & 5.87 & $-26.74$ & 1b & $J142200.24+001103.1$ & 5.89 & $-22.79$ & 3  & $J232908.28-030158.8$ & 6.42 & $-25.37$ & 2a \\
$J085813.52+000057.1$ & 5.99 & $-25.28$ & 3  & $J142517.72-001540.8$ & 6.18 & $-23.44$ & 3  & $J232914.46-040324.1$ & 5.90 & $-24.26$ & 2a \\
$J085907.19+002255.9$ & 6.39 & $-24.09$ & 3  & $J142920.23-000207.5$ & 6.04 & $-23.42$ & 3  & $J235651.58+002333.3$ & 6.00 & $-24.84$ & 1c \\
$J091833.17+013923.4$ & 6.19 & $-23.71$ & 3  & $J150941.78-174926.8$ & 6.12 & $-26.93$ & 2a &  & & &
\enddata
\tablecomments{The survey codes (SC) represent the SDSS main (1a), SDSS overlap (1b), SDSS stripe 82 (1c), CFHQS wide (2a), CFHQS deep (2b), and SHELLQs (3) surveys.
A full description of the individual objects may be found in \citet{jiang16} for the SDSS quasars, in \citet{willott10} for the CFHQS quasars, and in our previous papers for the SHELLQs quasars.
$J231546.58-002357.9$ was also recovered by the CFHQS and SHELLQs, and is hence included in the complete samples of all the three surveys. 
Five quasars in the SHELLQs sample ($J021013.19-045620.8$, $J022743.29-060530.3$, $J121721.34+013142.6$,  $J220417.92+011144.8$, $J221917.22+010249.0$) were originally discovered 
by other surveys \citep[see Table 1 of][for the details]{paperIV},
but are not included in the SDSS or CFHQS complete sample.}
\end{deluxetable*}

\begin{deluxetable*}{ccrrrrrrr}
\tablecaption{Number of objects in the $M_{1450}$ bins  \label{tab:LFcalc}}
\tabletypesize{\scriptsize}
\tablehead{
\colhead{$M_{1450}$} &  \colhead{$\Delta M_{1450}$} & \colhead{SDSS-main} & \colhead{SDSS-overlap}  &
\colhead{SDSS-S82} &  \colhead{CFHQS-W} & \colhead{CFHQS-D} & \colhead{SHELLQs} & \colhead{Total}
} 
\startdata
  $-22.00$  & 1.0  & 0 (  0.000) &   0 (  0.000) &   0 (  0.000) &   0 (  0.000)  &  1 (  0.003)  &  0  (0.058)  & 1  (0.062)   \\
  $-22.75$  & 0.5 &  0 (  0.000) &   0 (  0.000) &   0 (  0.000) &   0 (  0.000) &   0 (  0.014) &   8  (0.681)  &  8  (0.694)   \\
  $-23.25$  & 0.5 &  0 (  0.000) &   0 (  0.000) &   0 (  0.000)  &  0 (  0.000)  &  0 (  0.020) &   9  (1.629)  & 9 (1.649)   \\
  $-23.75$  & 0.5 &  0 (  0.000) &   0 (  0.000) &   0 (  0.000) &   0 (  0.072) &   0 (  0.023) &  10  (2.307)  & 10  (2.403) \\
  $-24.25$  & 0.5 &  0 (  0.000)  &  0 (  0.000) &   1 (  0.179) &   2 (  0.494) &   0 (  0.024) &   8   (2.645)  & 11  (3.341) \\
  $-24.75$  & 0.5 &  0 (  0.000) &   0 (  0.000) &   4 (  0.791) &   6 (  1.207) &   0 (  0.024) &   7  (2.811)  & 17  (4.833)  \\
  $-25.25$  & 0.5 &  0 (  0.000)  &  0 (  0.000) &   3 (  1.322) &   5 (  1.883) &   0 (  0.024) &   6  (2.911)  & 14  (6.140) \\
  $-25.75$  & 0.5 &  0 (  0.000)  &  1 (  0.619) &   3 (  1.606) &   1 (  2.282) &   0 (  0.024) &   0  (2.969)  & 5  (7.501) \\
  $-26.25$  & 0.5 &  1 (  3.647)  &  5 (  7.170) &   2 (  1.652) &   0 (  2.376) &   0 (  0.024) &   0  (3.005)  & 8  (17.874) \\
  $-26.75$  & 0.5 &  8 (25.859) &   2 (  8.251) &   0 (  1.645) &   2 (  2.355) &   0 (  0.024) &   0  (3.025)  & 12  (41.159) \\
  $-27.50$  & 1.0 & 14 (56.040)  &  2 (  2.940) &   0 (  1.645) &   0 (  2.311) &   0 (  0.024) &   0  (3.040)  & 16  (66.002) \\
  $-29.00$  & 2.0 &  1 (56.040)  &  0 (  0.000)  &  0 (  1.645) &   0 (  2.311) &   0 (  0.024) &   0 (3.040)    &  1  (63.061) \\
\hline
Total   & 8.0 & 24 (141.587) & 10 (18.981) & 13 (10.485) & 16 (15.291) & 1 (0.255) & 48 (28.120) & 112\tablenotemark{*} (214.719)\\
\enddata
\tablenotetext{*}{The number of unique objects is 110; $J231546.58-002357.9$ ($M_{1450} = -25.41$) is included in SDSS-S82, CFHQS-W, and SHELLQs, and thus is triply counted (see the text).
}
\tablecomments{$M_{1450}$ and $\Delta M_{1450}$ represent the center and width of each magnitude bin, respectively.
	The numbers in the parentheses represent the cosmic volumes contained in the individual surveys ($V_a$; see Equation \ref{eq:va}), given in Gpc$^3$.}
\end{deluxetable*}

\subsection{SDSS} \label{subsec:sample_sdss}

We exploit a complete sample of 47 SDSS quasars at $5.7 \le z \le 6.5$, presented in \citet{jiang16}.
Of these, 24 quasars with $z_{\rm AB} \le 20$ mag were discovered in the SDSS main survey, using single-epoch imaging data with 54-sec exposures.
17 quasars (in which 7 quasars were also found in the main survey) with $20 \le z_{\rm AB} \le 20.5$ mag were discovered in the SDSS overlap regions, 
where two or more exposures were taken, due to the scanning strategy and repeated observations of some fields in the main survey.
The remaining 13 quasars with  $z_{\rm AB} \le 22$ mag were discovered in the SDSS Stripe 82 on the celestial equator, which was repeatedly scanned 70 -- 90 times \citep{annis14, jiang14}.
In total, these 47 quasars span the magnitude range from $M_{1450} = -30$ to $-24$ mag.
The absolute magnitudes ($M_{1450}$) were estimated by extrapolating the continuum spectrum redward of Ly$\alpha$ to rest-frame 1450 \AA, by assuming a power-law shape $f_{\lambda} \propto \lambda^{-1.5}$
(except for a few quasars, whose observed spectra covered that rest-frame wavelength, or whose near-IR spectra provided estimates of the continuum slope).
The effective area of the main, overlap, and Stripe 82 surveys are 11,240, 4223, and 277 deg$^2$, respectively. 

The selection completeness was estimated with model quasars, which were created using spectral simulations presented in \citet{mcgreer13}.
The models were designed to reproduce the observed colors of $\sim$60,000 quasars at $2.5 < z < 3.5$ in the SDSS Baryon 
Oscillation Spectroscopic Survey \citep{ross12}, and
took into account the observed relations between spectral features and luminosity, such as the Baldwin effect.
The effect of IGM absorption was modeled using the prescription of \citet{worseck11} extended to higher redshifts with the data from \citet{songaila10}, 
and was checked against the measurements of \citet{songaila04} and \citet{fan06}.
The electronic data of the completeness functions of each of the three surveys were kindly provided by Linhua Jiang in private communication.

\subsection{CFHQS} \label{subsec:sample_cfhqs}

We use a complete sample of 17 CFHQS quasars at $5.7 \le z \le 6.5$, presented in \citet{willott10}.
Of these, 12 quasars were discovered in the Red-sequence Cluster Survey 2 (RCS-2) field observed with the MegaCam on CFHT, with exposure times
of 500 and 360 sec in the $i$ and $z$ band, respectively.
Four quasars were discovered in the CFHTLS Very Wide (VW) field, imaged for 540 and 420 sec in the MegaCam $i$ and $z$ band, respectively.
These 16 quasars (``CFHQS-wide quasars", hereafter) span the magnitude range from $M_{1450} = -27$ to $-24$ mag.
The remaining quasar, with $M_{1450} = -22.2$ mag, was discovered in the CFHQS deep field, which is a combination of the CFHTLS Deep and 
the Subaru {\it XMM-Newton} Deep Survey (SXDS) fields.
The effective areas of the CFHQS wide (RCS-2 $+$ CFHTLS-VW) and deep (CFHTLS-Deep $+$ SXDS) fields are 494 and 4.47 deg$^2$, respectively.
The selection completeness was estimated with quasar models created from the observed spectra of 180 SDSS quasars at $3.1 < z < 3.2$.
The effect of IGM absorption was incorporated based on the data taken from \citet{songaila04}.
The electronic data of the completeness functions were kindly provided by Chris Willott in private communication.

The absolute magnitudes ($M_{1450}$) of the CFHQS quasars were originally estimated from the observed $J$-band fluxes with a template quasar spectrum.
For consistency with the measurements in the SDSS and the SHELLQs, we re-measured their $M_{1450}$ by extrapolating the continuum spectrum redward of Ly$\alpha$, 
assuming a power-law shape $f_{\lambda} \propto \lambda^{-1.5}$.
The resultant $M_{1450}$ values differ from the original (CFHQS) values by $-0.4 - +0.2$ mag for all but one quasar;
the exception is the faintest quasar $J021627.81-045534.1$, 
for which the new measurement indicates 0.7-mag fainter continuum luminosity than in the original measurement.
This quasar has an unusually strong Ly$\alpha$ line, contributing about 70 \% of the observed $z$-band flux \citep{willott09}.
It has a similar $z - J$ color to other high-$z$ quasars despite the strong contribution of Ly $\alpha$ to the $z$-band flux, suggesting that the $J$-band also has significant 
contribution from strong lines like \ion{C}{4} $\lambda$1549. 
If so, the continuum flux is significantly fainter than the $J$-band magnitude would indicate.

\subsection{SHELLQs} \label{subsec:sample_shellqs}

We use 48 SHELLQs quasars at $5.7 \le z \le 6.5$, discovered from the HSC-SSP Wide survey fields. 
HSC is a wide-field camera mounted on the Subaru Telescope \citep{miyazaki18}.
It has a nearly circular field of view of 1$^\circ$.5 diameter, covered 
by 116 2K $\times$ 4K fully depleted Hamamatsu CCDs, with a pixel scale of 0\arcsec.17.
The HSC-SSP survey \citep{aihara18_survey} has three layers with different combinations of area and depth.
The Wide layer is observing 1400 deg$^2$ in several discrete fields mostly along the celestial equator, with 
5$\sigma$ point-source depths of ($g_{\rm AB}$, $r_{\rm AB}$, $i_{\rm AB}$, $z_{\rm AB}$, $y_{\rm AB}$) = (26.5, 26.1, 25.9, 25.1, 24.4) mag measured in 2\arcsec.0 apertures.
The total exposure times range from 10 minutes in the $g$- and $r$-bands to 20 minutes in the $i$-, $z$-, and $y$-bands, divided into individual exposures of $\sim$3 minutes each.
The Deep and the UltraDeep layers are observing smaller areas (27 and 3.5 deg$^2$) down to deeper limiting magnitudes ($r_{\rm AB}$ = 27.1 and 27.7 mag, respectively).
Data reduction was performed with the dedicated pipeline {\it hscPipe} \citep{bosch18}. 
We use the point spread function (PSF) magnitude ($m_{\rm PSF, AB}$, or simply $m_{\rm AB}$) and the CModel magnitude ($m_{\rm CModel, AB}$), 
which are measured by fitting the PSF models and two-component, PSF-convolved galaxy models to the source profile, respectively \citep{abazajian04, bosch18}.
We utilize forced photometry, which measures source flux with a consistent aperture in all bands. 
The aperture is usually defined in the $z$ band for $i$-band dropout sources, including high-redshift quasars.
A full description of the HSC-SSP survey may be found in \citet{aihara18_survey}. 

The SHELLQs quasars used in this work were drawn from the HSC-SSP Wide survey fields.
While the candidate selection procedure has changed slightly through the course of the survey, we defined a single set of criteria to select the 48 objects.
We first queried the ``S17A" internal data release (containing all the data taken before 2017 May) of the SSP survey, with the following conditions:
\begin{eqnarray}
z_{\rm AB} < 24.5\ \&\ \sigma_z < 0.155\ \&\ i_{\rm AB} - z_{\rm AB} > 2.0 \nonumber \\
 \&\ z_{\rm AB} - z_{\rm CModel, AB} < 0.15\nonumber \\
 \&\ {\tt merge.peak.(g, r, z, y)}\ = ({\tt f}, {\tt f}, {\tt t}, {\tt t})\nonumber \\
 \&\ {\tt (z, y).inputcount.value} \ge (2, 2)\nonumber \\
 \&\ {\tt (i, z, y).pixelflags.edge} =  ({\tt f}, {\tt f}, {\tt f})\nonumber \\
 \&\ {\tt (i, z, y).pixelflags.saturatedcenter} = ({\tt f}, {\tt f}, {\tt f})\nonumber \\
 \&\ {\tt (i, z, y).pixelflags.crcenter} =  ({\tt f}, {\tt f}, {\tt f})\nonumber \\
 \&\ {\tt (i, z, y).pixelflags.bad} = ({\tt f}, {\tt f}, {\tt f})\nonumber \\
 \&\ {\tt (i, z, y).pixelflags.bright.objectcenter} = ({\tt f}, {\tt f}, {\tt f})\nonumber \\
\label{eq:query1}
\end{eqnarray}
The first line defines the selection limits of magnitude, photometry S/N, and color, while the second line rejects apparently extended objects 
\citep[see][and the following section]{paperI}. 
The {\tt merge.peak} flag is true ({\tt t}) if the source is detected in the specified band, and false  ({\tt f}) if not.
The quasars in the present complete sample are required to be observed in the $i$, $z$, and $y$ bands (but not necessarily in the $g$ or $r$ band), and to be detected both in the $z$ and $y$ bands.
The condition on the {\tt inputcount.value} flag requires that the query is performed on the fields where two or more exposures were taken in each of the $z$ and $y$ bands.
The last five conditions reject sources on the pixels that are close to the CCD edge, saturated, affected by cosmic rays, registered as bad pixels, or close to bright objects, in any of the $i$, $z$, or $y$ bands.

The sources selected above were matched, within 1\arcsec.0, to near-IR sources from the UKIDSS \citep{lawrence07} and Visible and Infrared Survey Telescope for Astronomy (VISTA) 
Kilo-degree Infrared Galaxy (VIKING) surveys \citep{edge13}.
We then calculated a Bayesian probability ($P_{\rm Q}^{\rm B}$) for each candidate being a quasar rather than a Galactic brown dwarf (BD), 
based on models for the spectral energy distribution (SED) and surface density as a function of magnitude \citep[see][for the details]{paperI}.
Our algorithm does not include galaxy models at present. 
We 
consider those sources with $P_{\rm Q}^{\rm B} > 0.1$ in the list of candidates for spectroscopy.
Only $\sim$10 \% of the final SHELLQs quasars have near-IR counterparts in practice, and they would have been selected as candidates with the HSC photometry alone;
the near-IR photometry is mainly used to reject contaminating BDs, which have much redder near-IR - optical colors than do high-$z$ quasars.

Finally, the candidates went through a screening process using the HSC images.
We first used an automatic algorithm with Source Extractor \citep{bertin96}, to remove apparently spurious sources (e.g., cosmic rays, transient objects, and CCD artifacts).
The algorithm rejects those sources whose photometry (in all the available bands) is not consistent within 
5$\sigma$ error between the stacked and individual pre-stacked images, and those sources whose shapes are too compact, diffuse, or elliptical to be celestial point sources.
We checked a portion of the rejected sources, and confirmed that no real, stable sources were rejected in this automatic procedure.
Indeed, we adopted conservative rejection criteria here, so that any ambiguous cases were passed through to the next stage.
The remaining candidates were then screened by eye, which removed additional problematic objects (mostly cosmic rays and transient sources).
The automatic procedure rejected $>$95 \% of the input candidates, and $\sim$80 \% of the remaining candidates were removed by eye.

The final spectroscopic identification is still underway, but now has been completed down to a limiting magnitude of $z_{\rm AB}^{\rm splim} \simeq 24.0$ mag.
The actual $z_{\rm AB}^{\rm splim}$ values vary from field to field, depending on the available telescope time when the individual fields were observable, and are summarized in Table \ref{tab:fields}.
In total, 48 quasars with $z_{\rm AB} \le z_{\rm AB}^{\rm splim}$ and spectroscopic redshifts $5.7 \le z \le 6.5$ were selected as the complete sample for the present work.
The remaining SHELLQs quasars were not in the sample because they are fainter than $z_{\rm AB}^{\rm splim}$, outside the above redshift range, or fail to meet one or more of the criteria
listed in Equation \ref{eq:query1}.
The absolute magnitudes ($M_{1450}$) were estimated in the same way as used for the SDSS quasars (see above).

\begin{deluxetable*}{cccccr}
\tablecaption{SHELLQs survey fields \label{tab:fields}}
\tablehead{
\colhead{Name} &\colhead{R.A. range} &\colhead{Decl. range} &  \colhead{Area} & \colhead{$z_{\rm AB}^{\rm splim}$} & \colhead{$N_{\rm obj}$} \\
\colhead{} &\colhead{(deg)} &\colhead{(deg)} &  \colhead{(deg$^2$)} & \colhead{(mag)} & \colhead{} 
} 
\startdata
  XMM               & 28 -- 41  & $-7$ -- $+3$           & 83.7 & 24.1 & 5\\
  GAMA09H      & 127 -- 155  & $-3$ -- $+6$       & 165.1 & 23.8 & 8\\
  WIDE12H       & 173 -- 200  & $-3$ -- $+3$       & 106.5 & 23.8 & 10\\
  GAMA15H      & 205 -- 227 & $-3$ -- $+3$        & 100.7 & 24.0 & 8\\
  VVDS              & 330 -- 357  & $-2$ -- $+7$       & 124.7 & 24.2 & 13\\
  HECTOMAP   & 220 -- 252  & $+42$ -- $+45$   & 65.4 & 24.0 & 4\\
  \hline
   Total & \nodata & \nodata & 646.1 & \nodata & 48\\
\enddata
\tablecomments{The field names refer to the distinct areas covered in the HSC-SSP survey to date; see \citet{aihara18_survey} for details.
$z_{\rm AB}^{\rm splim}$ and $N_{\rm obj}$ represent the spectroscopic limiting magnitude and the number of quasars
included in the present complete sample, respectively.
}
\end{deluxetable*}

The effective survey area 
was estimated with a random source catalog stored in the HSC-SSP database 
\citep{coupon18}.
The random points are placed over the entire survey fields, with surface density of 100 arcmin$^{-2}$, and each point contains the survey information at the corresponding position
(number of exposures, variance of background sky, pixel quality flags, etc.) for each filter.
We queried this random catalog with the pixel flag conditions presented in Equation \ref{eq:query1}.
The number of output points were then divided by the input surface density, 
giving the effective survey area as listed in Table \ref{tab:fields}.

The SDSS, CFHQS, and SHELLQs samples contain one quasar in common ($J231546.58-002357.9$).
This quasar is treated as an independent object in each of the individual survey volumes, in order not to underestimate the number density.

\section{SHELLQ{\scriptsize s} Completeness} \label{sec:completeness}

The SHELLQs quasar selection is known to be fairly complete at bright magnitudes, to which past wide-field surveys (such as SDSS and CFHQS) 
were sensitive.
The HSC-SSP S17A survey footprint contains 8 previously-known high-$z$ quasars with $i_{\rm AB} - z_{\rm AB} > 2.0$, and our selection 
recovered 7 of them. 
The remaining quasar is blended 
with a foreground galaxy, which boosted the $i$-band flux of the quasar measured by the HSC pipeline and caused it to be rejected.
We evaluate the actual selection completeness in this section.

\subsection{Source Detection} \label{subsec:detection}

\begin{figure}
\epsscale{1.2}
\plotone{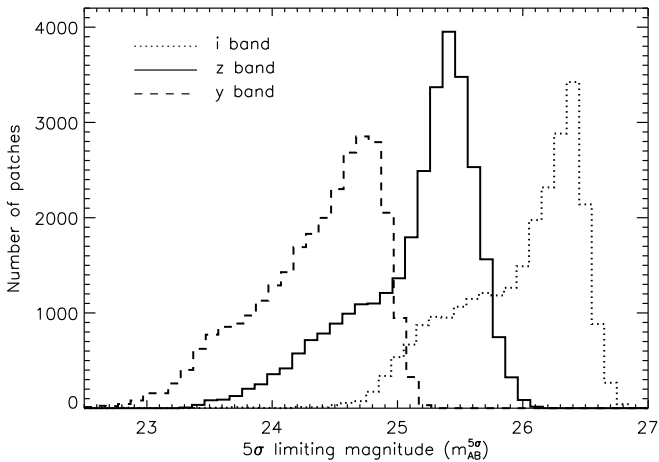}
\caption{Histograms of the 5$\sigma$ limiting magnitudes ($m_{\rm AB}^{5\sigma}$) measured in the 12\arcmin $\times$ 12\arcmin\ patches of the survey fields, 
in the $i$ (dotted), $z$ (solid) and $y$ (dashed) bands.
\label{fig:5sigmamag}}
\end{figure}

Source detection in the HSC data processing pipeline \citep[{\it hscPipe};][]{bosch18} is performed on PSF-convolved images, by finding pixels with flux $>$5$\sigma$ above the background sky.
Here $\sigma$ is the root-mean-square (RMS) of the local background fluctuations.
For a point source, this thresholding is approximately equivalent to $m_{\rm AB} < m_{\rm AB}^{5\sigma}$, where $m_{\rm AB}^{5\sigma}$ represents 
the PSF limiting magnitude at which S/N = 5
\citep[see][for a description of the theory]{bosch18}.
The HSC database stores $m_{\rm AB}^{5\sigma}$ measurements for each patch (12\arcmin $\times$ 12\arcmin) in the survey. 
As shown in Figure \ref{fig:5sigmamag}, 
all but a small fraction of the survey patches have $z_{\rm AB}^{5\sigma} > 24$ mag.
The $z$-band detection completeness is thus expected to be close to 100 \% for the quasars in our complete sample, which are brighter than $z_{\rm AB}^{\rm splim}$ = 23.8 -- 24.2 mag.

We tested the detection completeness in each band with simulations, 
in which artificial point sources were inserted on random positions of the stacked HSC images, and then recovered with {\it hscPipe}.
The input source models were created with the PSFs measured at each image position.
The same simulations were used in \citet{aihara18_pdr1} to evaluate the detection completeness of the HSC-SSP Public Data Release 1.\footnote{
More thorough simulations are possible with the {\it SynPipe} code \citep{huang18}, which we didn't use in the present work.}
These simulations were performed on 180 12\arcmin $\times$ 12\arcmin\ patches selected randomly from the survey area (the computer time required to run over 
the entire survey area would have been prohibitively long).
The recovery rate of the input sources, as a function of magnitude, 
is then fitted with a function \citep{serjeant00}:
\begin{equation}
f (m_{\rm AB}) = \frac{f_{\rm max} - f_{\rm min}}{2} ( \tanh [\alpha (m_{\rm AB}^{50} - m_{\rm AB})] + 1) + f_{\rm min}
\label{eq:completeness}
\end{equation}
where $f_{\rm max}$, $f_{\rm min}$, $\alpha$, and $m_{\rm AB}^{50}$ represent the detection completeness at the brightest and faintest magnitudes, the sharpness of the transition between 
$f_{\rm max}$ and $f_{\rm min}$, and the magnitude at which the detection completeness is 50 \%, respectively.

\begin{figure}
\epsscale{1.2}
\plotone{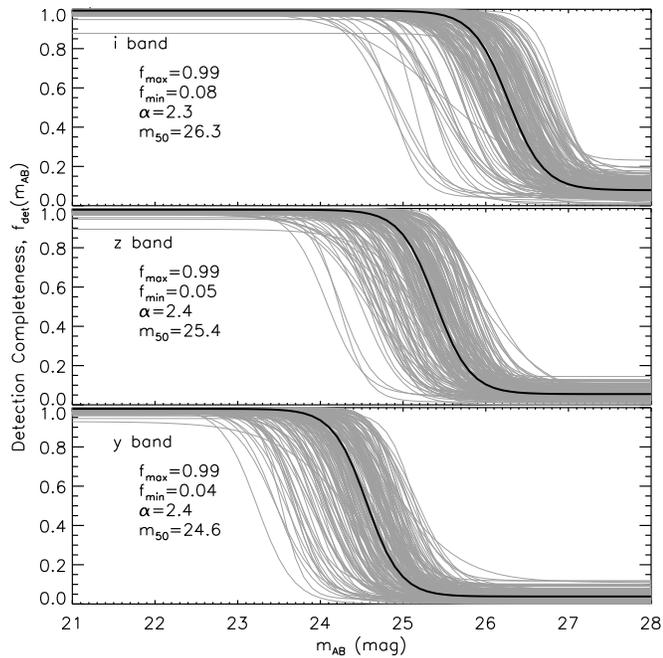}
\caption{Detection completeness 
in the $i$ (top), $z$ (middle), and $y$ (bottom) bands as modeled by Equation \ref{eq:completeness}, measured in each of the 180 random survey patches 
(thin gray lines).
The thick solid lines represent the median completeness, calculated with the median parameter values 
as reported in each panel.
\label{fig:detec_completeness}}
\end{figure}

\begin{figure}
\epsscale{1.2}
\plotone{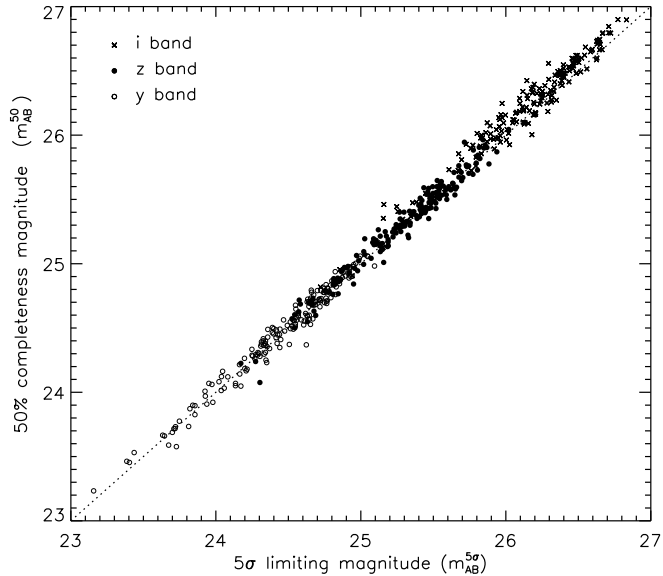}
\caption{Comparison between $m_{\rm AB}^{5\sigma}$ (5$\sigma$ limiting magnitudes) and $m_{\rm AB}^{50}$ (50-\% completeness magnitudes) in the $i$ (crosses), $z$ (dots), and $y$ (open circles) bands.
The dotted line represents $m_{\rm AB}^{5\sigma} = m_{\rm AB}^{50}$.
\label{fig:5sigma_m50}}
\end{figure}

The resultant completeness functions are presented in Figure \ref{fig:detec_completeness}. 
Overall they have similar shapes to each other, except for varying depths from patch to patch.
It is worth noting that the completeness at the faintest magnitudes ($f_{\rm min}$) is higher than zero, which is due to chance superposition of input sources with true sources 
in the original HSC images used. 
Figure \ref{fig:5sigma_m50} compares the $m_{\rm AB}^{50}$ values with the 5$\sigma$ limiting magnitudes ($m_{\rm AB}^{5\sigma}$) 
described above.
These two quantities agree very well with each other, as expected given that the {\it hscPipe} detection threshold is approximately equivalent to $m_{\rm AB} < m_{\rm AB}^{5\sigma}$.


Based on the above measurements and simulations, we quantified the detection completeness in the $z$ and $y$ bands over the entire survey area, as follows.
For each 12\arcmin $\times$ 12\arcmin\ patch (``$p$"), the completeness functions $f_{\rm det} (z_{\rm AB}, p)$ and $f_{\rm det} (y_{\rm AB}, p)$ were defined using Equation \ref{eq:completeness}.
We retrieved $z_{\rm AB}^{5\sigma}$ and $y_{\rm AB}^{5\sigma}$ from the survey database, and used them as surrogates for 
$z_{\rm AB}^{50}$ and $y_{\rm AB}^{50}$ in the individual patches.
The parameters $f_{\rm max}$ and $f_{\rm min}$ were fixed to 1.0 and 0.0, respectively. 
Finally we assumed $\alpha$ = 2.4, the median value measured in both the $z$ and $y$ bands for the 180 patches in which we ran the simulations 
(the dispersion in this quantity measured by the median absolute deviation is $\Delta\alpha \sim 0.4$ in both bands).
We checked that the present results are not sensitive to the choice of $\alpha$, since the detection completeness is close to 100 \% at the present magnitude limit of $z_{\rm AB} < 24.2$ mag.

\subsection{Point Source Selection} \label{subsec:pointsource}

\begin{figure}
\epsscale{1.2}
\plotone{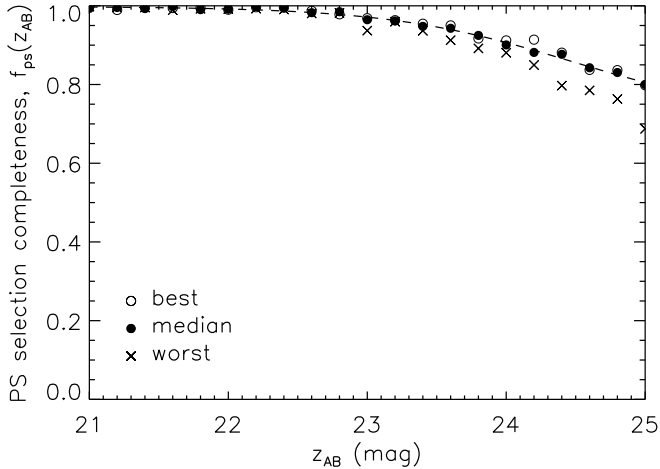}
\caption{Selection completeness of point sources, $f_{\rm ps} (z_{\rm AB})$, estimated with the {\it HST} ACS stars  
	on the SSP Wide-depth dataset of the COSMOS field.
	The open circles, dots, and crosses represent the best, median, and worst seeing conditions, respectively.
	The best-fit function (Equation \ref{eq:completeness}) to the median seeing data is represented by the dashed curve. 
\label{fig:pointsource}}
\end{figure}

The SHELLQs algorithm uses the criterion:
\begin{eqnarray}
z_{\rm AB} - z_{\rm CModel, AB} < 0.15
\label{eq:pointsource}
\end{eqnarray}
to identify point sources from the HSC database.
The completeness of this selection was evaluated with a special HSC dataset on the COSMOS field, one of the two UltraDeep fields of the SSP survey, for which we have many more
exposures than in a Wide field.
This dataset was created by stacking a portion of the UltraDeep data taken during the best, median, or worst seeing conditions to match the Wide depth.
We selected stars on this field with the {\it Hubble Space Telescope} ({\it HST}) Advanced Camera for Surveys (ACS) catalog \citep{leauthaud07},
and measured the fraction of stars meeting Equation \ref{eq:pointsource}.
The results are presented in Figure \ref{fig:pointsource}.
The completeness of our point source selection is close to 100 \% at bright magnitudes, and decreases mildly to 90 \% at $z_{\rm AB} \sim 24.0$ mag.
No significant difference was observed between the different seeing conditions at $z_{\rm AB} < 24$ mag.
We fitted the above results for the median seeing with Equation \ref{eq:completeness},
and obtained the best-fit parameters ($f_{\rm max}$, $f_{\rm min}$, $\alpha$, $z_{\rm AB}^{50}$) = (1.00, 0.72, 0.76, 24.5).
This best-fit function, $f_{\rm ps} (z_{\rm AB})$, is used to simulate the selection completeness of point sources in the following.

On the other hand, we found that the effect of resolved host galaxies on our quasar selection is negligible.
This was simulated as follows.
Since the luminosities of high-$z$ quasar host galaxies are unknown, we assumed the following, based on the low-$z$ results for SDSS quasars
with similar nuclear luminosity to the SHELLQs quasars \citep{matsuoka14, matsuoka15}: 
(i) the typical host galaxy luminosity ranges from $M_{UV} = -18$ to $-21$ mag (corresponding to $z_{\rm CModel, AB} \sim 25.5 - 28.5$ mag at $z = 6$), and 
(ii) there is no correlation between the nuclear and host galaxy luminosities.
The host galaxies were simulated with a sample of Lyman Break Galaxies (LBGs) at $z \sim 6$, found from the HSC-SSP Wide data \citep{harikane18, ono18}.
We used 231 LBGs with $24.0 < z_{\rm CModel, AB} < 25.0$ mag, where AGN contamination to the sample is small \citep{ono18}.
For each LBG, 
we randomly assigned $M_{UV}$ from $-18$ to $-21$ mag, 
assumed a flat UV spectral slope \citep[$\beta = -2.0$;][]{stanway05}, 
and calculated the corresponding CModel flux ($f_{\rm CModel}^{\rm sim}$) at $z = 6$.
The PSF flux was calculated as $f_{\rm PSF}^{\rm sim} = f_{\rm CModel}^{\rm sim}  \times (f_{\rm PSF}^{\rm obs} / f_{\rm CModel}^{\rm obs})$, where $f_{\rm PSF}^{\rm obs} / f_{\rm CModel}^{\rm obs}$
is the ratio between the PSF and CModel fluxes observed for the individual LBGs.
Then we added various AGN fluxes ($f^{\rm AGN} = f_{\rm PSF}^{\rm AGN} = f_{\rm CModel}^{\rm AGN}$) artificially, and calculated the fraction of 
the simulated objects that satisfy Equation \ref{eq:pointsource} and are thus ``unresolved":
\begin{equation}
-2.5 \log \left( \frac{f_{\rm PSF}^{\rm sim} + f^{\rm AGN}}{f_{\rm CModel}^{\rm sim} + f^{\rm AGN}} \right) < 0.15 .
\end{equation}
We found that the unresolved fraction is 100 \% at AGN magnitudes $z_{\rm AB} < 25.0$ mag, and decreases to 90 \% at 26.0 mag.
We thus conclude that our point source selection loses only a negligible fraction of quasars due to the resolved host galaxies,
at the present magnitude limit of 
$z_{\rm AB} < 24.2$ mag.

Here we note that compact galaxies could have $z_{\rm AB} - z_{\rm CModel, AB} < 0.15$, and contaminate our quasar candidates. 
Indeed, so far we have discovered 25 high-$z$ galaxies in addition to 74 high-$z$ quasars from the HSC candidates. 
However, the present work uses only spectroscopically confirmed quasars, and thus is not affected by galaxy contamination.

\subsection{Foreground flux contamination} \label{subsec:fgd}

As we wrote previously, we failed to recover one of the eight previously-known quasars in our survey footprint, due to $i$-band flux contamination of a foreground galaxy.
The forced photometry can overestimate the $i$-band flux of an $i$-band dropout object superposed on a foreground source, because the aperture is 
defined by the object image in a redder band.

\begin{figure}
\epsscale{1.2}
\plotone{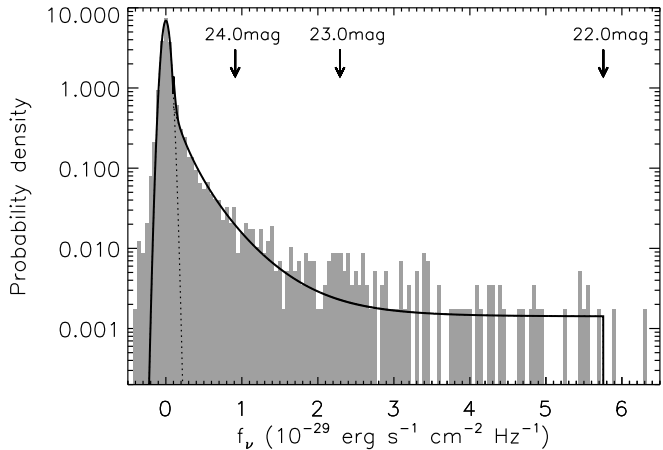}
\caption{Probability density distribution of the $i$-band fluxes measured on random positions (histogram).
The solid line represents the best-fit function, which is a combination of a Gaussian function (dotted line) and the function $f_{\rm fgd} (f_\nu)$ defined in the text.
The arrows mark the fluxes corresponding to $i_{\rm AB} = 22.0$, 23.0, and 24.0 mag.}
\label{fig:random_apphoto}
\end{figure}

In order to simulate this effect, we randomly selected 10000 points from the HSC-SSP random source catalog in the way that we described in \S \ref{subsec:sample_shellqs},
and measured the $i$-band flux in an aperture placed at each point.
The aperture size was set to twice the seeing FWHM at each position.
The probability density distribution (PDF) of the measured fluxes is presented in Figure \ref{fig:random_apphoto}.
The distribution around $f_{\nu} = 0$ follows a Gaussian distribution, which represents the sky background fluctuation.
In addition, the measured distribution has a tail toward higher $f_{\nu}$, which can be approximated by the function\footnote{
This functional form was arbitrarily determined to fit the data.}
 $f_{\rm fgd} (f_\nu)\ =\ 3.3\ e^{-5\sqrt{f_{\nu, 29}}} + 0.0014$ ($f_{\nu, 29} = f_\nu \times 10^{29}$ erg s$^{-1}$ cm$^{-2}$ Hz$^{-1}$)
truncated at $f_{\nu, 29} = 5.8$ (corresponding to $i_{\rm AB} = 22.0$ mag, above which the measured PDF contains less than 0.5 \% of the total probability).
This tail contains 12 \% of the total probability, which is the fraction of sources affected by the foreground flux contamination.
We use this function $f_{\rm fgd} (f_\nu)$ in the following simulations.

The foreground flux contamination is much less significant in the $z$ and $y$ bands, in which high-$z$ quasars (meeting Equation \ref{eq:query1}) are clearly detected 
and the {\it hscPipe} deblender properly apportions the measured flux.
\citet{huang18} demonstrated that the HSC flux measurement is accurate within 0.1 mag after deblending for the vast majority of the sources.

\subsection{Total Completeness}

\begin{figure}
\epsscale{1.1}
\plotone{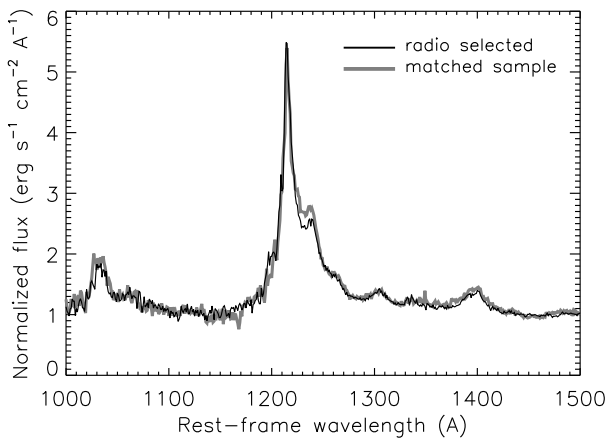}
\caption{Composite spectra of 29 radio-selected SDSS quasars (black dashed line) and of a matched sample of 29 quasars selected by other criteria (gray solid line) at $z \simeq 3$.
These composite spectra were created by converting the individual spectra to rest-frame wavelengths and normalizing the flux at 1450 \AA, and then averaging all the input spectra.}
\label{fig:radio_z3quasars}
\end{figure}

The total completeness of our selection was estimated with quasar models, created from 319 SDSS spectra of luminous ($-27 \le M_{i} \le -30$) quasars at $z \simeq 3$.
This SDSS sample contains 29 radio-selected quasars, which are not sensitive to incompleteness in the color selection \citep[see, e.g.,][]{worseck11}.
We selected a sample of 29 non-radio-selected quasars (i.e., objects selected for SDSS spectroscopy with other targeting criteria) from the remaining 290 objects, matched 
in luminosity to the radio-selected quasars, and compared the composite spectra of the two samples.
This is shown in Figure \ref{fig:radio_z3quasars}.
The composite spectra are almost identical to each other, 
indicating that the colors of radio- and color-selected quasars are similar, and that we introduce no significant bias by using the spectra of all 319 quasars in the simulations which follow.
We note that the above radio-selected quasars are still a part of the magnitude-limited SDSS sample, and are biased against optically-faint populations such as obscured quasars. 
The present estimate does not include incompleteness due to such quasars that are missing from the SDSS spectroscopic sample.

Each of the above 319 spectra was redshifted to $z$ = 5.6 -- 6.6, with $\Delta z = 0.01$ steps, with appropriate correction for the different amounts of IGM \ion{H}{1} absorption 
between $z \sim 3$ and $z \sim 6$.
The IGM absorption in the original SDSS spectra was removed using the mean IGM effective optical depth ($\tau_{\rm eff}$) at $z \le 3$ presented by \citet{songaila04}.
We then added IGM absorption to the redshifted model spectra by assuming the mean and scatter of $\tau_{\rm eff}$ taken from \citet{eilers18}.
The absorption started at a wavelength corresponding to 1 proper Mpc from the quasar, to model the effect of quasar proximity zones.  
The assumed proximity radius 
is appropriate for the mean luminosity
of the SHELLQs quasars \citep[$M_{1450} \sim -23$ mag;][]{eilers17}.
The damping wing of the IGM absorption was modeled following the prescription in \citet{totani06}.

At this stage we found that the mean and the scatter of rest-frame Ly$\alpha$ equivalent widths (EWs) of the model quasars were 64 $\pm$ 16 \AA\ (this includes the effect of IGM absorption, and was 
measured with a subset of model quasars matched in redshift to the observed sample;
the scatter was measured with the median absolute deviation), which are larger than those of the observed sample, 38 $\pm$ 12 \AA.
This trend is opposite to the luminosity dependence known as the Baldwin effect, and may be in part due to the redshift dependence of quasar SEDs, including a higher fraction 
of weak-line quasars found at higher redshifts \citep[e.g.,][]{banados16, shen18}.
We scaled the Ly$\alpha$ line of the model spectra, 
with the scaling factor chosen randomly from a Gaussian distribution of mean 0.6 and standard deviation 0.2, which roughly reproduces the observed EW distribution.
Since the HSC bands cover only a limited portion (rest-frame wavelength $\la 1500$ \AA) of the high-$z$ quasar spectra redward of Ly$\alpha$, differences
in other emission lines or continuum slopes between the $z \sim 3$ SDSS quasars and the SHELLQs quasars would not be very relevant here.

The simulations of our quasar selection were performed with 
five million points 
selected from the HSC-SSP random source catalog, using the pixel flag conditions in Equation \ref{eq:query1}.
We randomly assigned one of the above quasar models to each random point, and calculated apparent magnitudes, assuming an absolute magnitude drawn from
a uniform distribution from $M_{1450} = -20$ to $-28$ mag.
We then added simulated errors to the apparent magnitudes, assuming a Gaussian error distribution with standard deviation ($\sigma$) equal to the sky background RMS, 
computed from the $5\sigma$ limiting magnitudes of the corresponding patches ($m^{5\sigma}_{\rm AB}$; see above).
We simulated the foreground flux contamination using the PDF $f_{\rm fgd} (f_\nu)$, derived in \S \ref{subsec:fgd}.

\begin{figure*}
\epsscale{1}
\plotone{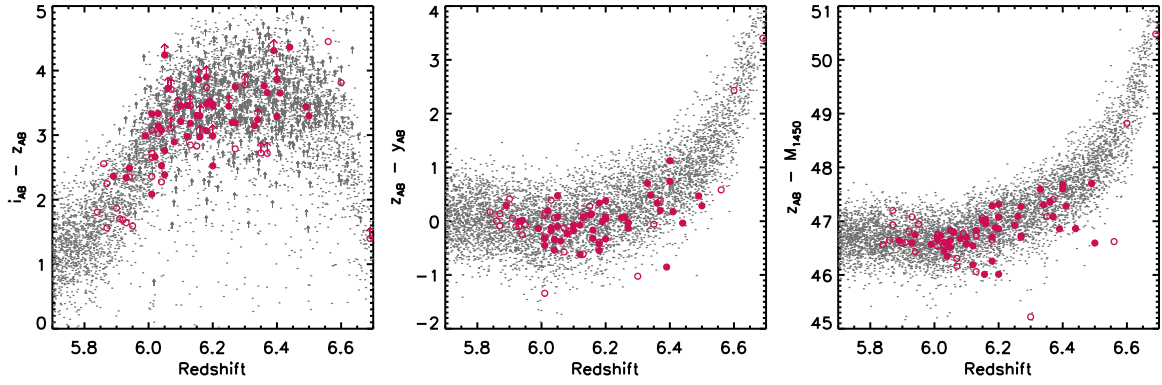}
\caption{The $i_{\rm AB} - z_{\rm AB}$ (left), $z_{\rm AB} - y_{\rm AB}$ (middle), and $z_{\rm AB} - M_{1450}$ (right) distributions of the simulated quasars with $z_{\rm AB} < 24.2$ mag (gray dots). 
The arrows represent 2$\sigma$ lower limits.
The SHELLQs quasars included in and excluded from the present complete sample are represented by the filled and open circles, respectively.}
\label{fig:color_distr}
\end{figure*}

We then applied additional flux scatter with a Gaussian distribution with standard deviation 0.3 mag, in each of the three bands.
This was necessary to match the color distributions of the model and observed quasars, while it does not change the derived LF significantly.
This additional scatter 
may account for other sources of flux fluctuation than explicitly considered above, including photometry errors due to cosmic rays, image artifacts, and imperfect source deblending,
the host galaxy contribution, and difference in the intrinsic SED shapes between the above SDSS quasars and the SHELLQs quasars
\citep[see, e.g.,][]{niida16}.
The resultant color distributions of the model and observed quasars are presented in Figure \ref{fig:color_distr}.

We selected a portion of the above simulated quasars, such that a quasar with simulated magnitudes ($z_{\rm AB}$, $y_{\rm AB}$) on a patch $p$ has a probability
$f_{\rm det} (z_{\rm AB}, p) \times f_{\rm det} (y_{\rm AB}, p) \times f_{\rm ps} (z_{\rm AB})$ of being selected.
This accounts for the field variance of the detection completeness.
We further selected those meeting the following conditions:
\begin{equation}
z_{\rm AB} < z_{\rm AB}^{\rm slim}\ \&\ \sigma_z < 0.155\ \&\ i_{\rm AB} - z_{\rm AB} > 2.0 .
\end{equation}
Finally we calculated Bayesian quasar probabilities ($P_{\rm Q}^{\rm B}$) for the selected sources, using the method described in \citet{paperI}, 
and counted the number of sources with $P_{\rm Q}^{\rm B} > 0.1$.
The total completeness, $f_{\rm comp} (z, M_{1450})$, is given by the ratio between the output and input numbers of random sources, calculated in bins of $z$ and $M_{1450}$.
There are roughly 400 simulated quasars in each bin with sizes $\Delta{z} = 0.01$ and $\Delta{M_{1450}} = 0.05$.

\begin{figure}
\epsscale{1.2}
\plotone{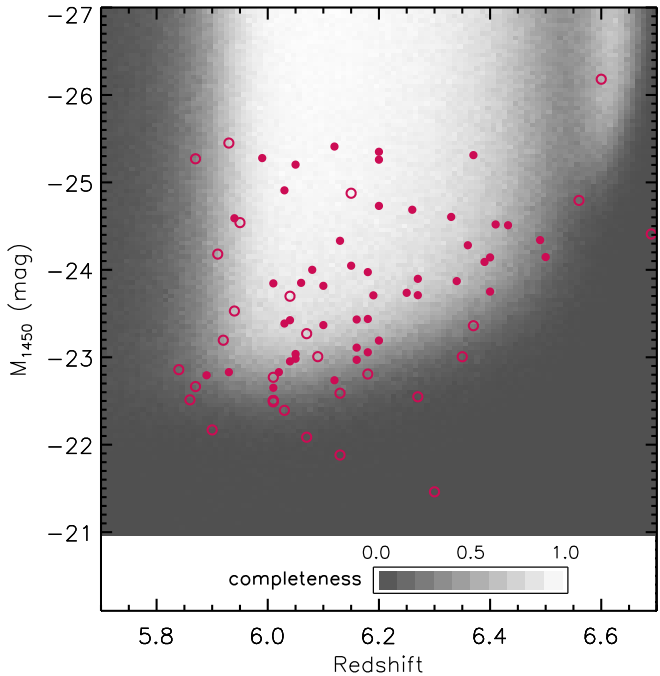}
\caption{Total completeness of the SHELLQs complete quasar selection, ranging from $f_{\rm comp} (z, M_{1450}) = 1.0$ in white to 0.0 in gray.
The SHELLQs quasars included in and excluded from the present complete sample are marked by the filled and open circles, respectively.
\label{fig:completeness}}
\end{figure}

Figure \ref{fig:completeness} presents the total completeness derived above.
The selection of the present complete sample 
is most sensitive to $5.9 < z < 6.5$ and $M_{1450} < -22.5$ mag. 
The completeness drops at $z \le 5.9$ due to the color cut of $i - z > 2.0$, while it drops more gradually at $z \ge 6.5$ due to the increasing contamination of brown dwarfs (which reduces the quasar probability $P_{\rm Q}^{\rm B}$).
The figure also shows that several quasars located in the high completeness region are not included in the complete sample.
This is caused by various reasons; some quasars are in survey fields that fail to meet the pixel flag conditions (Equation \ref{eq:query1}) in the S17A data release, and some quasars have $i - z$ colors just below the threshold of 2.0.
The faintest quasars with $M_{1450} > -22.5$ mag simply fail to meet the condition $z_{\rm AB} < z_{\rm AB}^{\rm slim}$.

In the following section, we use the completeness functions of the SDSS, CFHQS, and SHELLQs to derive a single LF.
These functions were all derived with quasar models tied to spectra of SDSS quasars at $z \sim 3$, while the IGM absorption models in the SDSS and CFHQS were created from older  $\tau_{\rm eff}$ data 
than those we used here for the SHELLQs sample.
We tested another IGM absorption model for the SHELLQs sample, with the mean and scatter of the $\tau_{\rm eff}$ determined empirically to reproduce the data in \citet{songaila04}, and found 
little change in the derived completeness or LF.
In addition, while the completeness correction is most important at the faintest luminosity of a given sample, the faintest SDSS/CFHQS quasars have smaller available volumes ($V_{\rm a}$; see below
and Table \ref{tab:LFcalc}) and thus smaller weights in LF calculation than do the CFHQS/SHELLQs quasars with similar luminosities and high completeness.
Thus we conclude that no significant bias is introduced by combining the completeness functions of the three surveys.

\section{Luminosity Function} \label{sec:LF}

\begin{figure}
\epsscale{1.2}
\plotone{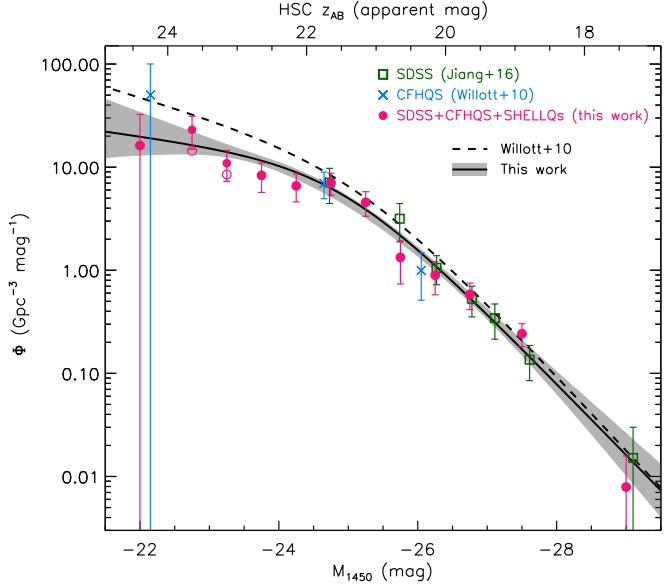}
\caption{Binned LF measured by the SDSS \citep[squares;][]{jiang16}, the CFHQS  \citep[crosses;][]{willott10}, and this work combining
the SDSS, CFHQS, and SHELLQs samples (dots).
The open circles show the LF excluding the five quasars with narrow Ly$\alpha$ (see the text).
The solid line represents our parametric LF with the 1$\sigma$ confidence interval shown by the shaded area,
while the dashed line represents the parametric LF of \citet{willott10}.
All the parametric LFs are calculated at $z = 6.0$.
\label{fig:LF}}
\end{figure}

First, we derive the binned LF using the 1/$V_{\rm a}$ method \citep{avni80}. 
The cosmic volume available to discover a quasar, in a magnitude bin $\Delta M_{1450}$, is given by
\begin{equation}
V_{\rm a} = \frac{1}{\Delta M_{1450}}   \int_{\Delta M_{1450}} \int_{\Delta z}  f_{\rm comp} (z, M_{1450})\ \frac{dV_{\rm c}}{dz}\ dz\ dM_{1450} ,
\label{eq:va}
\end{equation}
where $\Delta z$ represents the redshift range to calculate the LF, and $dV_{\rm c}/dz$ is the co-moving volume element probed by a survey.
The binned LF and its uncertainty are then given by
\begin{eqnarray}
\Phi_{\rm b} (M_{1450}) = \frac{1}{\Delta M_{1450}} \sum \frac{1}{V_{\rm a}} ,\nonumber\\
\Delta \Phi_{\rm b} (M_{1450}) = \frac{1}{\Delta M_{1450}} \left[ \sum \left( \frac{1}{V_{\rm a}} \right)^2 \right]^{1/2} ,
\label{eq:binnedLF}
\end{eqnarray}
where the sum is taken over the quasars in the magnitude bin.
This expression ignores the redshift evolution of the LF over the measured range ($5.7 \le z \le 6.5)$; we will take this evolution into account in the parametric LF described below.
Here we combine the three complete samples of quasars from the SDSS, the CFHQS, and the SHELLQs, to derive a single binned LF over $-22 < M_{1450} < -30$ mag
(we use the completeness functions and the survey areas of the SDSS and CFHQS described in \S \ref{subsec:sample_sdss} and \S \ref{subsec:sample_cfhqs}).
We set the bin size $\Delta M_{1450} = 0.5$ mag, except at both ends of the luminosity coverage where the sample size is small.
The results of this calculation are listed in Table \ref{tab:binnedLF} and presented in Figure \ref{fig:LF}.

The derived LF agrees well with the previous results from the SDSS \citep{jiang16} and the CFHQS \citep{willott10} at $M_{1450} < -25$ mag, and significantly improves the accuracy at fainter magnitudes.
It may be worth mentioning that the number density of the brightest bin measured by \citet{jiang16} and in this work do not exactly match, although the two works use a single SDSS quasar in common.
This is due to the different choice of the bin center and width, which is known to have a significant impact on the binned LF when the sample size is small.
On the other hand, we significantly increased the available survey volume for the faintest bin at $M_{1450} = -22.00$, and found a number density lower than (but consistent within 1$\sigma$) 
the previous measurement by \citet{willott10}.

\begin{deluxetable}{cccc}
\tablecaption{Binned luminosity function \label{tab:binnedLF}}
\tabletypesize{\scriptsize}
\tablehead{
\colhead{$M_{1450}$} &\colhead{$\Delta M_{1450}$} &\colhead{$\Phi_{\rm b} (M_{1450})$} &  \colhead{$N_{\rm obj}$}\\
\colhead{} &\colhead{} &\colhead{(Gpc$^{-3}$ mag$^{-1}$)} &  \colhead{}
}
\startdata
 $-22.00$ & 1.0 &   16.2 $\pm$ 16.2    & 1  \\
 $-22.75$ & 0.5 &    23.0 $\pm$  8.1   & 8  \\
 $-23.25$  & 0.5 & 10.9 $\pm$  3.6  & 9 \\
 $-23.75$ & 0.5 &  8.3 $\pm$  2.6    & 10 \\
 $-24.25$  & 0.5 &  6.6 $\pm$  2.0   & 11 \\
 $-24.75$  & 0.5 &  7.0 $\pm$  1.7   & 17 \\
 $-25.25$  & 0.5 &   4.6 $\pm$  1.2   & 14 \\
 $-25.75$  & 0.5 &  1.33 $\pm$  0.60  &  5\\
 $-26.25$  & 0.5 &   0.90  $\pm$ 0.32 &  8\\
 $-26.75$ & 0.5 &   0.58 $\pm$  0.17   &  12\\
 $-27.50$ & 1.0 &    0.242 $\pm$  0.061   &  16\\
 $-29.00$ & 2.0 &   0.0079 $\pm$  0.0079  & 1 \\
\hline\hline
 $-22.75$ & 0.5 &    14.4 $\pm$  6.4   & 5  \\
 $-23.25$  & 0.5 &   8.5 $\pm$  3.2    & 7 \\
\enddata
\tablecomments{$M_{1450}$ and $\Delta M_{1450}$ represent the center and width of each magnitude bin, respectively.
	$N_{\rm obj}$ represents the number of quasars contained in the bin.
	The last two rows report the LF at $-22.5 < M_{1450} < -23.5$ excluding narrow Ly$\alpha$ quasars (see the text).}
\end{deluxetable}

Next, we derive the parametric LF, using a commonly-used double power-law function:
\begin{equation}
\Phi_{\rm p} (M_{1450}) =  \frac{10^{k (z - 6)} \Phi^*} {10^{0.4 (\alpha + 1) (M_{1450} - M_{1450}^*)} + 10^{0.4 (\beta + 1) (M_{1450} - M_{1450}^*)}} ,
\end{equation}
where $\alpha$ and $\beta$ are the faint- and bright-end slopes, respectively.
We fix the redshift evolution term to $k = -0.47$ \citep{willott10} or $k = -0.7$ \citep{jiang16}; we found that the choice makes little difference in the determination of other parameters
(see below). 
Following the argument in \citet{jiang16}, we adopt $k = -0.7$ as our standard value.
The parameters $M_{1450}^*$ and $\Phi^*$ give the break magnitude and normalization of the LF, respectively.

\begin{figure*}
\epsscale{1.0}
\plotone{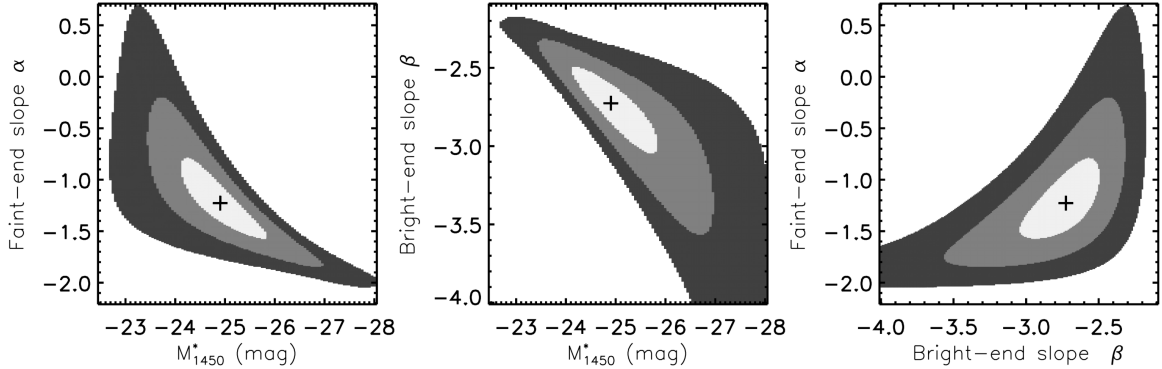}
\caption{Confidence regions (light gray: 1$\sigma$, gray: 2$\sigma$, dark gray: 3$\sigma$) of the individual LF parameters.
The best-fit values are marked by the crosses.
\label{fig:plf_par}}
\end{figure*}

We perform a maximum likelihood fit \citep{marshall83} to determine the four free parameters ($\alpha$, $\beta$, $M_{1450}^*$, and $\Phi^*$).
Specifically, we maximize the likelihood $L$ by minimizing $S = -2 \ln{L}$, given by
\begin{eqnarray}
S = -2 \sum \ln \left[ \Phi_{\rm p} (z, M_{1450})\ f_{\rm comp} (z, M_{1450}) \right] \nonumber\\
      + 2 \int_{-\infty}^{+\infty} \int_{-\infty}^{+\infty} \Phi_{\rm p} (M_{1450}, z)\ f_{\rm comp} (z, M_{1450})\ \frac{dV_{\rm c}}{dz}\ dz\ dM_{1450} , \nonumber\\
\end{eqnarray}
where the sum in the first term is taken over all quasars in the sample.
The resultant parametric LF is presented in Figure \ref{fig:LF}, and the best-fit LF parameters 
are listed in the first row of Table \ref{tab:parametricLF}.
Figure \ref{fig:plf_par} presents the confidence regions of the individual LF parameters.

\begin{deluxetable*}{lccccc}
\tablecaption{Parametric luminosity function \label{tab:parametricLF}}
\tabletypesize{\scriptsize}
\tablehead{
\colhead{} & \colhead{$\Phi^*$} &\colhead{$M^*_{1450}$} &\colhead{$\alpha$} &  \colhead{$\beta$} & \colhead{$k$}\\
\colhead{} & \colhead{(Gpc$^{-3}$ mag$^{-1}$)} &  \colhead{} & \colhead{} &\colhead{} &\colhead{}
}
\startdata
Standard       &                                           $10.9^{+10.0}_{-6.8}$ & $-24.90^{+0.75}_{-0.90}$ & $-1.23^{+0.44}_{-0.34}$  &  $-2.73^{+0.23}_{-0.31}$   & $-0.7$ \\\hline
Different $k$ &                                           $9.5^{+9.6}_{-6.2}$  & $-25.02^{+0.82}_{-0.98}$  & $-1.27^{+0.42}_{-0.33}$  & $-2.74^{+0.24}_{-0.33}$ & $-0.47$ \\
Free $k$                                                    & $7.8^{+9.2}_{-5.6}$  & $-25.18^{+0.88}_{-1.13}$  & $-1.34^{+0.43}_{-0.34}$  &  $-2.76^{+0.26}_{-0.40}$  &  $-0.2^{+0.2}_{-0.1}$ \\
Narrow-Ly$\alpha$ quasars excluded      & $14.1^{+6.8}_{-6.7}$   & $-24.64^{+0.54}_{-0.66}$ & $-0.88^{+0.48}_{-0.39}$  & $-2.67^{+0.18}_{-0.25}$  & $-0.7$ \\
Quasars with $z > 5.9$                             & $8.1^{+12.3}_{-5.9}$   & $-25.30^{+1.05}_{-1.15}$  &  $-1.39^{+0.45}_{-0.32}$ & $-2.79^{+0.32}_{-0.48}$ & $-0.7$ \\
\enddata
\end{deluxetable*}

This is the first time that observed data have shown a clear break in the LF for 
$z \sim 6$ quasars.
The bright-end slope, $\beta = -2.73^{+0.23}_{-0.31}$, agrees very well with those reported previously by \citet[][$\beta = -2.81$, with the faint-end slope fixed to $\alpha = -1.5$]{willott10}
and \citet[][$\beta = -2.8 \pm 0.2$, fitting only the brightest portion of the LF]{jiang16}.
The break magnitude is $M_{1450}^* = -24.90^{+0.75}_{-0.90}$, and the LF flattens significantly toward lower luminosities.
The slope $\alpha = -1.23^{+0.44}_{-0.34}$ is even consistent with a completely flat faint-end LF (i.e., $\alpha = 1.0$).

We also performed LF calculations with $k$ fixed to $-0.47$ or allowed to vary as a free parameter, and found that the other LF parameters are not very sensitive to the choice of $k$.
These results are listed in the second and third rows of Table \ref{tab:parametricLF}.
The fitting with variable $k$ favors relatively flat LF evolution ($k = -0.2^{+0.2}_{-0.1}$), which may be consistent with a tendency that 
$k$ is smaller for lower-luminosity quasars seen in \citet[][their Figure 10]{jiang16}.
But given the short redshift baseline of the present sample, we chose to adopt the fixed value $k = -0.7$ for our standard LF.

Recently, \citet{kulkarni18} reported a very bright break magnitude of $M_{1450}^* = -29.2^{+1.1}_{-1.9}$ mag at $z \sim 6$, by re-analyzing the quasar sample constructed by \citet{jiang16}, \citet{willott10}, and \citet{kashikawa15}.
However, their data favor a single power-law LF, and thus the break magnitude was forced to be at the bright end of the sample in their LF fitting \citep{kulkarni18}.
The present work indicates that the LF breaks at a much fainter magnitude, in the luminosity range that has been poorly explored previously.


It may be worth noting that the CFHQS-deep survey discovered one quasar in the $M_{1450} = -22.00$ bin from $V_{\rm a} = 0.003$ Gpc$^3$, while SHELLQs discovered no quasars (in the present complete sample) 
in the same $M_{1450}$ bin from $V_{\rm a} = 0.058$ Gpc$^3$ (Table \ref{tab:LFcalc}).
This is presumably due to 
statistical fluctuations.
Based on the present parametric LF, 
the expected total number of quasars in the CFHQS-deep survey is roughly one, with the most likely luminosity 
in the range $-25 \la M_{1450} \la -22$ mag.
In reality the survey discovered one quasar with $M_{1450} = -21.5$ mag and none at brighter magnitudes, which is consistent with the expectation.
On the other hand, the expected number of SHELLQs quasars in the $M_{1450} = -22.00$ bin is roughly one.
This is consistent with the actual discovery of no quasars in this bin, 
given Poisson noise.

The SHELLQs complete sample used here includes five objects with narrow Ly$\alpha$ lines (FWHM $< 500$ km s$^{-1}$) at $-23.5 < M_{1450} < -22.5$.
We classified them as quasars based on their extremely high Ly$\alpha$ luminosities, featureless continuum, and possible mini broad absorption line system of 
\ion{N}{5} $\lambda$1240 seen in their composite spectrum \citep{paperII}.
It is possible that they are not in fact type-1 quasars, so 
for reference, we re-calculated the binned LF at $-23.5 < M_{1450} < -22.5$ omitting these five objects, and listed the results in the last two rows of Table \ref{tab:binnedLF}.
The parametric LF in this case is reported in the fourth row of Table \ref{tab:parametricLF}, which shows a modest difference from the standard case.

We also calculated the LF by limiting the sample to the 89 quasars in our complete sample at $z > 5.9$, the redshift range over which the CFHQS and SHELLQs are 
most sensitive (see Figure \ref{fig:sample}).
The resultant parametric LF is listed in the last row of Table \ref{tab:parametricLF}.
The LF in this case has slightly brighter $M^*_{1450}$ and steeper $\alpha$ than the standard LF, but the difference is smaller than the fitting uncertainty.

\begin{figure}
\epsscale{1.2}
\plotone{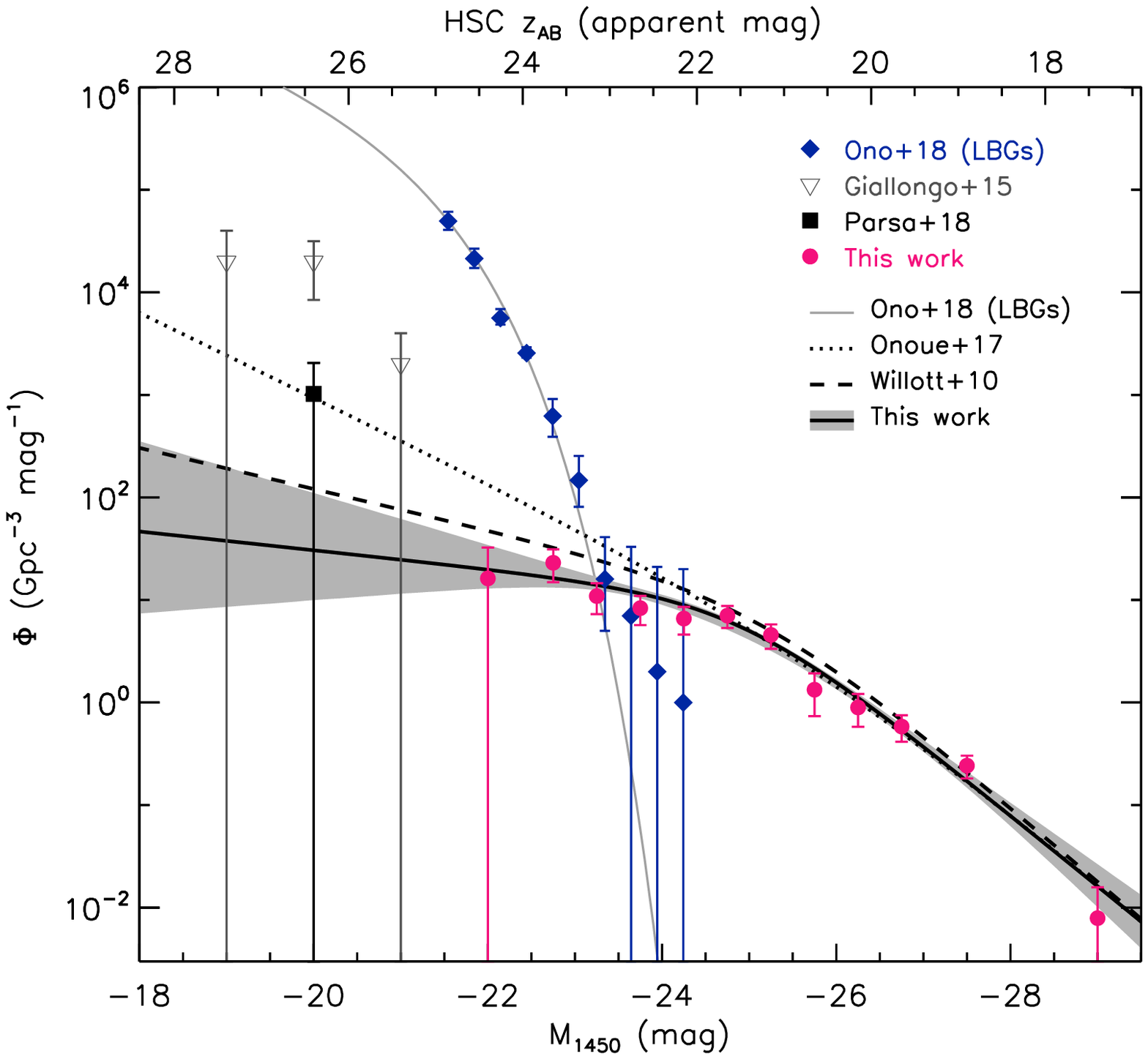}
\caption{Binned LFs measured by \citet[][for LBGs; diamonds]{ono18}, \citet[][triangles]{giallongo15}, \citet[][squares]{parsa18}, and this work (dots).
In the X-ray measurements by \citet{giallongo15} and \citet{parsa18}, the rest-UV magnitudes $M_{1450}$ were estimated from the optical photometry of the galaxies matched to the X-ray sources.
The lines represent the parametric LFs measured by \citet[][for LBGs; gray solid]{ono18}, \citet[][their case $1^\prime$; dotted]{onoue17}, \citet[][dashed]{willott10}, and this work (solid; the 1$\sigma$ confidence
interval is shown by the shaded area).
All the parametric LFs are calculated at $z = 6.0$.
\label{fig:LF2}}
\end{figure}

Figure \ref{fig:LF2} displays our LF and several past measurements below the break magnitude, $M_{1450} \ge -25$ mag.
We found a flatter LF than reported in \citet{willott10} and \citet[][and their previous paper \citet{kashikawa15}]{onoue17}, 
who had only a few low-luminosity quasars in their samples.
The extrapolation of our LF underpredicts the number densities of faint AGNs compared to those reported by \citet{giallongo15}, 
while the former is consistent with the more recent measurements by \citet{parsa18}.
On the other hand, we note that the above X-ray measurements are immune to dust obscuration, and that 
the discrepancy with the rest-UV measurements, if any, could be due to the presence of a large population of obscured AGNs in the high-$z$ universe.
Finally, Figure \ref{fig:LF2} indicates that LBGs \citep[taken from][]{ono18} outnumber quasars at $M_{1450} > -23$ mag.
This is consistent with our experience from the SHELLQs survey, 
which found increasing numbers of LBGs contaminating the quasar candidate sample at $z_{\rm AB} > 23$ mag \citep{paperI, paperII, paperIV}. 

\begin{figure}
\epsscale{1.1}
\plotone{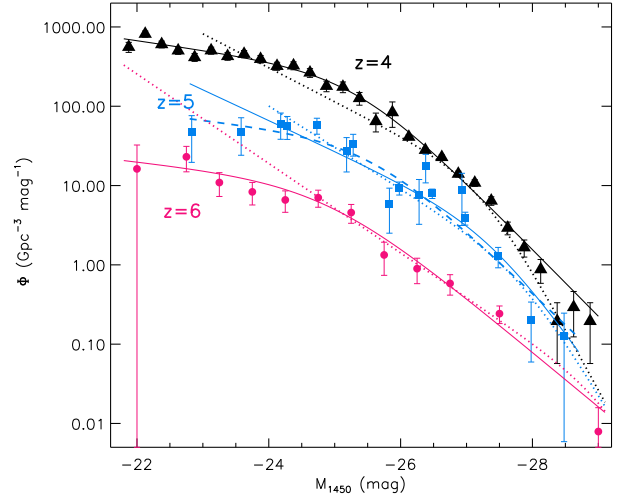}
\caption{Binned LFs at $z \sim 4$ \citep[triangles;][]{akiyama18}, $z \sim 5$ \citep[squares;][]{mcgreer18}, and $z \sim 6$ (dots; this work), along with the parametric LFs at those redshifts
(the three solid lines).
The dashed line represents the parametric LF fitted to the \citet{mcgreer18} data with the fixed bright-end slope $\beta = -3.0$ (see the text),
while the three dotted lines represent the parametric LFs at the three redshifts reported by \citet{kulkarni18}.
\label{fig:LF_evol}}
\end{figure}

We compare the present LF with those recently derived at $z \sim 4$ \citep{akiyama18} and $z \sim 5$ \citep{mcgreer18} in Figure \ref{fig:LF_evol}.
The overall shape of the binned LF remains relatively similar, while there is a steep decline of the total number density toward higher redshifts.
However, the best-fit break magnitudes reported in the above studies differ substantially, i.e., $M_{1450}^*$ = ($-25.36 \pm0.13$, $-27.42^{+0.22}_{-0.26}$, $-24.90^{+0.75}_{-0.90}$)
at $z \sim$ (4, 5, 6).
This may be in part due to the choice of the fixed bright-end slope $\beta = -4.0$ in \citet{mcgreer18}, which is significantly steeper than measured at $z \sim 4$ \citep[$\beta \sim -3.1$;][]{akiyama18}
or at $z \sim 6$ ($\beta \sim -2.7$; this work).
As shown in the middle panel of Figure \ref{fig:plf_par}, the bright-end slope and the break magnitude are strongly covariant in the parametric LF fitting.
We found that the binned LF of \citet{mcgreer18} can also be fitted reasonably well with $\beta = -3.0$, as shown in Figure \ref{fig:LF_evol} (dashed line).
The best-fit break magnitude in this case is $M_{1450}^* = -25.6 \pm 0.3$, which is close to the break magnitudes at $z \sim 4$ and $z \sim 6$.
The figure also displays the parametric LFs reported by \citet{kulkarni18};
while these LFs match the data in the luminosity ranges covered by their sample, the LFs seem to overpredict the number densities of fainter quasars presented in the recent studies
by \citet{akiyama18}, \citet{mcgreer18}, and this paper.


Since the LF is a product of the mass function and the Eddington ratio function of SMBHs, it is not straightforward to interpret the significant flattening
observed at $M_{1450} \ge -25$ mag, in terms of a unique physical model.
It could indicate relatively mass-independent number densities and/or quasar radiation efficiency, at low SMBH masses.
We will compare our LF with theoretical models in a forthcoming paper.
Alternatively, as discussed above, the LF flattening may indicate an increasing fraction of obscured AGNs toward low luminosities,
especially in light of the X-ray results in Figure \ref{fig:LF2}.
This could be an interesting subject for future deep X-ray observations, such as those that {\it ATHENA} \citep{nandra13} will achieve.

\section{Contribution to Cosmic Reionization} \label{sec:reionization}

There is much debate about the source of photons that are responsible for cosmic reionization, as we discussed in \S \ref{sec:intro}.
Here we derive the total ionizing photon density from quasars per unit time, $\dot{n}_{\rm ion}$ (s$^{-1}$ Mpc$^{-3}$), and compare with that necessary to keep the IGM fully ionized.
The ionizing photon density can be calculated as:
\begin{equation}
\dot{n}_{\rm ion}= f_{\rm esc}\ \epsilon_{1450}\ \xi_{\rm ion} ,
\end{equation}
where $f_{\rm esc}$ is the photon escape fraction, $\epsilon_{1450}$  (erg s$^{-1}$ Hz$^{-1}$ Mpc$^{-3}$) is the total photon energy density from quasars at 1450 \AA:
\begin{equation}
\epsilon_{1450} = \int \Phi_{\rm p} (M_{1450}, z) L_{1450} dM_{1450} ,
\label{eq:epsilon}
\end{equation}
and $\xi_{\rm ion}$ [s$^{-1}$/(erg s$^{-1}$ Hz$^{-1}$)] is the number of ionizing photons from a quasar with a monochromatic luminosity
$L_{1450}$ = 1 erg s$^{-1}$ Hz$^{-1}$ at 1450 \AA:
\begin{equation}
\xi_{\rm ion} = (L_{1450})^{-1} \int^{4\nu_{\rm LL}}_{\nu_{\rm LL}} \frac{L_\nu}{h\nu} d\nu .
\label{eq:xi}
\end{equation}
Equation \ref{eq:epsilon} was integrated from $M_{1450} = -18$ to $-30$ mag, using the parametric LF derived in the previous section.
In Equation \ref{eq:xi}, we used a broken power-law quasar SED ($f_\nu \propto \nu^{-1.70}$ at $\lambda < 912\ $\AA\ and $\propto \nu^{-0.61}$ at $\lambda > 912\ $\AA) presented by \citet{lusso15}, 
and integrated from the \ion{H}{1} Lyman limit (frequency $\nu = \nu_{\rm LL}$) to the \ion{He}{2} Lyman limit ($\nu = 4 \nu_{\rm LL}$).
The implicit assumptions here are that the above SED, created from luminous quasars at $z \sim 2.4$, holds for the present high-$z$ quasars, and that 
all the ionizing photons with $\nu < 4 \nu_{\rm LL}$ are absorbed by the IGM.
The resultant photon density is 
$\dot{n}_{\rm ion} = 10^{48.8 \pm 0.1}$ s$^{-1}$ Mpc$^{-3}$ at $z = 6.0$ for $f_{\rm esc} = 1$.
We would get lower $\dot{n}_{\rm ion}$ for $f_{\rm esc} < 1$, which may be the case for low-luminosity quasars \citep{cristiani16, micheva17, grazian18}.
The energy density at 912 \AA\ is estimated to be $\epsilon_{912} = 10^{22.9 \pm 0.1}$ erg s$^{-1}$ Hz$^{-1}$ Mpc$^{-3}$, which is close to the value 
reported by \citet{haardt12} at $z = 6$.
The results presented in this section change very little when the faint limit of the integral in Equation \ref{eq:epsilon} is changed to $M_{1450} = -10$ mag, or when the five SHELLQs 
quasars with narrow Ly $\alpha$ (see \S \ref{sec:LF}) are excluded.

On the other hand, the evolution of the \ion{H}{2} volume-filling factor in the IGM, $Q_{\rm H II} (t)$, is given by
\begin{equation}
\frac{dQ_{\rm H II}}{dt} = \frac{\dot{n}_{\rm ion}}{\bar{n}_{\rm H}} - \frac{Q_{\rm H II}}{\bar{t}_{\rm rec}} ,
\label{eq:Q_HII}
\end{equation}
where $\bar{n}_{\rm H}$ and $\bar{t}_{\rm rec}$ are the mean hydrogen density and recombination time, respectively \citep{madau99}.
In the ionized IGM with $Q_{\rm H II} = 1.0$, the rate of ionizing photon density which balances recombination is given by
\begin{equation}
\dot{n}^{\rm crit}_{\rm ion} = \frac{\bar{n}_{\rm H}}{\bar{t}_{\rm rec}} = 10^{50.0} C_{\rm H II} \left( \frac{1+z}{7} \right)^3 ({\rm s}^{-1} {\rm Mpc}^{-3}) ,
\label{eq:Nion_crit}
\end{equation}
where $C_{\rm H II}$ represents an effective \ion{H}{2} clumping factor \citep{bolton07}.
The ionizing photon density we found above, given our LF, is less than 10 \% of 
$\dot{n}^{\rm crit}_{\rm ion}$ 
for the plausible range of $C_{\rm H II} = 1.0 - 5.0$ \citep{shull12}.
This means that quasars alone cannot sustain reionization.
For reference, we would get $\dot{n}_{\rm ion} = 10^{50.3}$ s$^{-1}$ Mpc$^{-3}$ $\sim \dot{n}^{\rm crit}_{\rm ion}$
if we assumed no LF break ($\alpha = \beta = -2.73$) and integrated Equation \ref{eq:epsilon} from $M_{1450} = -18$ to $-30$ mag.

\begin{figure}
\epsscale{1.2}
\plotone{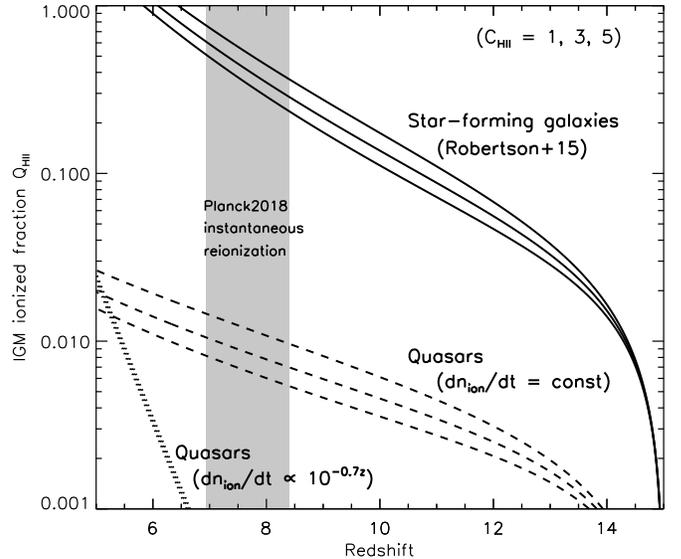}
\caption{Evolution of the \ion{H}{2} volume-filling factor in the IGM. 
The three solid curves represent contribution from star-forming galaxies \citep{robertson15} for the clumping factor $C_{\rm H II}$ = 1, 3, 5 from top to bottom.
The dashed and dotted curves represent the quasar contribution for the same $C_{\rm H II}$ values, for models with constant $\dot{n}_{\rm ion}$ or $\dot{n}_{\rm ion} \propto 10^{-0.7z}$, respectively (see the text).
The shaded area represents the 1$\sigma$ confidence interval of the instantaneous reionization redshift, taken from the {\it Planck} measurements \citep{planck18}.
\label{fig:Q_HII}}
\end{figure}

Finally, we numerically integrate Equation \ref{eq:Q_HII} and track the evolution of $Q_{\rm H II}$ driven solely by quasar radiation.
We assume that the IGM was neutral at $z = 15$, and that $\dot{n}_{\rm ion}$ was constant in time (i.e., it stayed at $10^{48.8 \pm 0.1}$ s$^{-1}$ Mpc$^{-3}$)
or evolved as $\propto 10^{-0.7z}$ (i.e., proportional to the LF normalization found around $z = 6$) at $5 < z < 15$.
We followed \citet{robertson15} to estimate $\bar{n}_{\rm H}$ and $\bar{t}_{\rm rec}$.
The results of this calculation are presented in Figure \ref{fig:Q_HII}. 
For reference, we also plot the $Q_{\rm H II}$ evolution driven by star-forming galaxies, using the star formation rate density at $z < 15$ presented in \citet{robertson15}.
This figure demonstrates that star-forming galaxies can supply enough high-energy photons to ionize the IGM by $z = 6$, while quasars cannot.
We thus conclude that quasars are not a major contributor to reionization.
Even if there is a large population of obscured AGNs that are missed by rest-UV surveys (see the discussion in \S \ref{sec:LF}), they are unlikely to release many 
ionizing photons, since the ionizing photon escape fraction from these objects would be close to $f_{\rm esc} \sim 0$.

\section{Summary \label{sec:summary}}

This paper presented new measurements of the quasar LF at $z \sim 6$, which is now established over an unprecedentedly wide magnitude range from $M_{1450} = -30$ to $-22$ mag.
We collected a complete sample of 110 quasars from the SDSS, the CFHQS, and the SHELLQs surveys.
The completeness of the SHELLQs quasar selection was carefully evaluated, and we showed that the selection is most sensitive to quasars with $5.9 < z < 6.5$ and $M_{1450} < -22.5$ mag.
The resultant binned LF is consistent with previous results at $M_{1450} < -25$ mag, while it exhibits significant flattening at fainter magnitudes.
The maximum likelihood fit of a double power-law function to the sample yielded a faint-end slope $\alpha =  -1.23^{+0.44}_{-0.34}$, a bright-end slope $\beta = -2.73^{+0.23}_{-0.31}$, 
a break magnitude $M_{1450}^* = -24.90^{+0.75}_{-0.90}$, and a characteristic space density $\Phi^* = 10.9^{+10.0}_{-6.8}$ Gpc$^{-3}$ mag$^{-1}$.
The rate of ionizing photon density from quasars is
$\dot{n}_{\rm ion} = 10^{48.8 \pm 0.1}$ s$^{-1}$ Mpc$^{-3}$, when integrated over $-18 < M_{1450} < -30$ mag.
This accounts for $<$10 \% of the critical rate necessary to keep the IGM fully ionized at $z = 6.0$.
We conclude that quasars are not a major contributor to cosmic reionization.

The HSC-SSP survey is making steady progress toward its goal of observing 1,400 deg$^2$ in the Wide layer.
We will continue follow-up spectroscopy to construct a larger complete sample of $z \sim 6$ quasars, down to lower luminosity than probed in the present work.
We are also starting an intensive effort to explore higher redshifts, with the aim of establishing the quasar LF at $z \sim 7$.
At the same time, we are collecting near-IR spectra to measure the SMBH masses and mass accretion rates, which will be used in combination with the LFs
to understand the growth of SMBHs in the early Universe.
The ALMA follow-up observations are also ongoing, which will provide valuable information on the formation and evolution of the host galaxies.

\acknowledgments
This work is based on data collected at the Subaru Telescope and retrieved from the HSC data archive system, which is operated by the Subaru Telescope and Astronomy Data Center at 
National Astronomical Observatory of Japan (NAOJ).
The data analysis was in part carried out on the open use data analysis computer system at the Astronomy Data Center of NAOJ.


YM was supported by the Japan Society for the Promotion of Science (JSPS) KAKENHI Grant No. JP17H04830 and the Mitsubishi Foundation Grant No. 30140.
NK acknowledges supports from the JSPS grant 15H03645.
KI acknowledges support by the Spanish MINECO under grant AYA2016-76012-C3-1-P and MDM-2014-0369 of ICCUB (Unidad de Excelencia 'Mar\'ia de Maeztu').

The Hyper Suprime-Cam (HSC) collaboration includes the astronomical communities of Japan and Taiwan, and Princeton University.  The HSC instrumentation and software were developed by the NAOJ, the Kavli Institute for the Physics and Mathematics of the Universe (Kavli IPMU), the University of Tokyo, the High Energy Accelerator Research Organization (KEK), the Academia Sinica Institute for Astronomy and Astrophysics in Taiwan (ASIAA), and Princeton University.  Funding was contributed by the FIRST program from Japanese Cabinet Office, the Ministry of Education, Culture, Sports, Science and Technology (MEXT), the Japan Society for the Promotion of Science (JSPS),  Japan Science and Technology Agency  (JST),  the Toray Science  Foundation, NAOJ, Kavli IPMU, KEK, ASIAA,  and Princeton University.

The Pan-STARRS1 Surveys (PS1) have been made possible through contributions of the Institute for Astronomy, the University of Hawaii, the Pan-STARRS Project Office, the Max-Planck Society and its participating institutes, the Max Planck Institute for Astronomy, Heidelberg and the Max Planck Institute for Extraterrestrial Physics, Garching, The Johns Hopkins University, Durham University, the University of Edinburgh, Queen's University Belfast, the Harvard-Smithsonian Center for Astrophysics, the Las Cumbres Observatory Global Telescope Network Incorporated, the National Central University of Taiwan, the Space Telescope Science Institute, the National Aeronautics and Space Administration under Grant No. NNX08AR22G issued through the Planetary Science Division of the NASA Science Mission Directorate, the National Science Foundation under Grant No. AST-1238877, the University of Maryland, and Eotvos Lorand University (ELTE).
 
This paper makes use of software developed for the Large Synoptic Survey Telescope. We thank the LSST Project for making their code available as free software at http://dm.lsst.org.


\begin{thebibliography}{}
\bibitem[Abazajian et al.(2004)]{abazajian04} Abazajian, K., Adelman-McCarthy, J.~K., Ag{\"u}eros, M.~A., et al.\ 2004, \aj, 128, 502 
\bibitem[Aihara et al.(2018a)]{aihara18_pdr1} Aihara, H., Armstrong, R., Bickerton, S., et al.\ 2018, \pasj, 70, S8 
\bibitem[Aihara et al.(2018b)]{aihara18_survey} Aihara, H., Arimoto, N., Armstrong, R., et al.\ 2018, \pasj, 70, S4 
\bibitem[Akiyama et al.(2018)]{akiyama18} Akiyama, M., He, W., Ikeda, H., et al.\ 2018, \pasj, 70, S34 
\bibitem[Annis et al.(2014)]{annis14} Annis, J., Soares-Santos, M., Strauss, M.~A., et al.\ 2014, \apj, 794, 120 
\bibitem[Avni \& Bahcall(1980)]{avni80} Avni, Y., \& Bahcall, J.~N.\ 1980, \apj, 235, 694 
\bibitem[Ba{\~n}ados et al.(2016)]{banados16} Ba{\~n}ados, E., Venemans, B.~P., Decarli, R., et al.\ 2016, \apjs, 227, 11 
\bibitem[Ba{\~n}ados et al.(2018)]{banados18} Ba{\~n}ados, E., Venemans, B.~P., Mazzucchelli, C., et al.\ 2018, \nat, 553, 473 
\bibitem[Bertin \& Arnouts(1996)]{bertin96} Bertin, E., \& Arnouts, S.\ 1996, \aaps, 117, 393 
\bibitem[Bolton \& Haehnelt(2007)]{bolton07} Bolton, J.~S., \& Haehnelt, M.~G.\ 2007, \mnras, 382, 325 
\bibitem[Bosch et al.(2018)]{bosch18} Bosch, J., Armstrong, R., Bickerton, S., et al.\ 2018, \pasj, 70, S5 
\bibitem[Bouwens et al.(2011)]{bouwens11} Bouwens, R.~J., Illingworth, G.~D., Labbe, I., et al.\ 2011, \nat, 469, 504 
\bibitem[Bouwens et al.(2015)]{bouwens15} Bouwens, R.~J., Illingworth, G.~D., Oesch, P.~A., et al.\ 2015, \apj, 803, 34 
\bibitem[Cappelluti et al.(2016)]{cappelluti16} Cappelluti, N., Comastri, A., Fontana, A., et al.\ 2016, \apj, 823, 95 
\bibitem[Chambers et al.(2016)]{chambers16} Chambers, K.~C., Magnier, E.~A., Metcalfe, N., et al.\ 2016, arXiv:1612.05560 
\bibitem[Coupon et al.(2018)]{coupon18} Coupon, J., Czakon, N., Bosch, J., et al.\ 2018, \pasj, 70, S7 
\bibitem[Cristiani et al.(2016)]{cristiani16} Cristiani, S., Serrano, L.~M., Fontanot, F., Vanzella, E., \& Monaco, P.\ 2016, \mnras, 462, 2478 
\bibitem[D'Aloisio et al.(2017)]{daloisio17} D'Aloisio, A., Upton Sanderbeck, P.~R., McQuinn, M., Trac, H., \& Shapiro, P.~R.\ 2017, \mnras, 468, 4691 
\bibitem[Edge et al.(2013)]{edge13} Edge, A., Sutherland, W., Kuijken, K., et al.\ 2013, The Messenger, 154, 32 
\bibitem[Eilers et al.(2018)]{eilers18} Eilers, A.-C., Davies, F.~B., \& Hennawi, J.~F.\ 2018, \apj, 864, 53 
\bibitem[Eilers et al.(2017)]{eilers17} Eilers, A.-C., Davies, F.~B., Hennawi, J.~F., et al.\ 2017, \apj, 840, 24 
\bibitem[Fan et al.(2006)]{fan06} Fan, X., Strauss, M.~A., Becker, R.~H., et al.\ 2006, \aj, 132, 117 
\bibitem[Fukugita et al.(1996)]{fukugita96} Fukugita, M., Ichikawa, T., Gunn, J.~E., et al.\ 1996, \aj, 111, 1748 
\bibitem[Giallongo et al.(2015)]{giallongo15} Giallongo, E., Grazian, A., Fiore, F., et al.\ 2015, \aap, 578, A83 
\bibitem[Giavalisco et al.(2004)]{giavalisco04} Giavalisco, M., Ferguson, H.~C., Koekemoer, A.~M., et al.\ 2004, \apjl, 600, L93 
\bibitem[Glikman et al.(2011)]{glikman11} Glikman, E., Djorgovski, S.~G., Stern, D., et al.\ 2011, \apjl, 728, L26 
\bibitem[Grazian et al.(2018)]{grazian18} Grazian, A., Giallongo, E., Boutsia, K., et al.\ 2018, \aap, 613, A44 
\bibitem[Haardt \& Madau(2012)]{haardt12} Haardt, F., \& Madau, P.\ 2012, \apj, 746, 125 
\bibitem[Harikane et al.(2018)]{harikane18} Harikane, Y., Ouchi, M., Ono, Y., et al.\ 2018, \pasj, 70, S11 
\bibitem[Huang et al.(2018)]{huang18} Huang, S., Leauthaud, A., Murata, R., et al.\ 2018, \pasj, 70, S6 
\bibitem[Ikeda et al.(2012)]{ikeda12} Ikeda, H., Nagao, T., Matsuoka, K., et al.\ 2012, \apj, 756, 160 
\bibitem[Ikeda et al.(2011)]{ikeda11} Ikeda, H., Nagao, T., Matsuoka, K., et al.\ 2011, \apjl, 728, L25 
\bibitem[Ishigaki et al.(2018)]{ishigaki18} Ishigaki, M., Kawamata, R., Ouchi, M., et al.\ 2018, \apj, 854, 73 
\bibitem[Izumi et al.(2018)]{izumi18} Izumi, T., Onoue, M., Shirakata, H., et al.\ 2018, arXiv:1802.05742 (Paper III)
\bibitem[Jiang et al.(2014)]{jiang14} Jiang, L., Fan, X., Bian, F., et al.\ 2014, \apjs, 213, 12 
\bibitem[Jiang et al.(2016)]{jiang16} Jiang, L., McGreer, I.~D., Fan, X., et al.\ 2016, \apj, 833, 222 
\bibitem[Kashikawa et al.(2015)]{kashikawa15} Kashikawa, N., Ishizaki, Y., Willott, C.~J., et al.\ 2015, \apj, 798, 28 
\bibitem[Khaire(2017)]{khaire17} Khaire, V.\ 2017, \mnras, 471, 255 
\bibitem[Kulkarni et al.(2018)]{kulkarni18} Kulkarni, G., Worseck, G., \& Hennawi, J.~F.\ 2018, arXiv:1807.09774 
\bibitem[Lawrence et al.(2007)]{lawrence07} Lawrence, A., Warren, S.~J., Almaini, O., et al.\ 2007, \mnras, 379, 1599 
\bibitem[Leauthaud et al.(2007)]{leauthaud07} Leauthaud, A., Massey, R., Kneib, J.-P., et al.\ 2007, \apjs, 172, 219 
\bibitem[Lusso et al.(2015)]{lusso15} Lusso, E., Worseck, G., Hennawi, J.~F., et al.\ 2015, \mnras, 449, 4204 
\bibitem[Madau et al.(1999)]{madau99} Madau, P., Haardt, F., \& Rees, M.~J.\ 1999, \apj, 514, 648 
\bibitem[Marshall et al.(1983)]{marshall83} Marshall, H.~L., Tananbaum, H., Avni, Y., \& Zamorani, G.\ 1983, \apj, 269, 35 
\bibitem[Masters et al.(2012)]{masters12} Masters, D., Capak, P., Salvato, M., et al.\ 2012, \apj, 755, 169 
\bibitem[Matsuoka et al.(2018b)]{paperIV} Matsuoka, Y., Iwasawa, K., Onoue, M., et al.\ 2018b, \apjs, 237, 5
\bibitem[Matsuoka et al.(2016)]{paperI} Matsuoka, Y., Onoue, M., Kashikawa, N., et al.\ 2016, \apj, 828, 26
\bibitem[Matsuoka et al.(2018a)]{paperII} Matsuoka, Y., Onoue, M., Kashikawa, N., et al.\ 2018a, \pasj, 70, S35
\bibitem[Matsuoka et al.(2015)]{matsuoka15} Matsuoka, Y., Strauss, M.~A., Shen, Y., et al.\ 2015, \apj, 811, 91 
\bibitem[Matsuoka et al.(2014)]{matsuoka14} Matsuoka, Y., Strauss, M.~A., Price, T.~N., III, \& DiDonato, M.~S.\ 2014, \apj, 780, 162 
\bibitem[McGreer et al.(2018)]{mcgreer18} McGreer, I.~D., Fan, X., Jiang, L., \& Cai, Z.\ 2018, \aj, 155, 131 
\bibitem[McGreer et al.(2013)]{mcgreer13} McGreer, I.~D., Jiang, L., Fan, X., et al.\ 2013, \apj, 768, 105 
\bibitem[McLeod et al.(2016)]{mcleod16} McLeod, D.~J., McLure, R.~J., \& Dunlop, J.~S.\ 2016, \mnras, 459, 3812 
\bibitem[Micheva et al.(2017)]{micheva17} Micheva, G., Iwata, I., \& Inoue, A.~K.\ 2017, \mnras, 465, 302 
\bibitem[Mitra et al.(2018)]{mitra18} Mitra, S., Choudhury, T.~R., \& Ferrara, A.\ 2018, \mnras, 473, 1416 
\bibitem[Miyazaki et al.(2018)]{miyazaki18} Miyazaki, S., Komiyama, Y., Kawanomoto, S., et al.\ 2018, \pasj, 70, S1 
\bibitem[Miyazaki et al.(2002)]{miyazaki02} Miyazaki, S., Komiyama, Y., Sekiguchi, M., et al.\ 2002, \pasj, 54, 833 
\bibitem[Mortlock et al.(2011)]{mortlock11} Mortlock, D.~J., Warren, S.~J., Venemans, B.~P., et al.\ 2011, \nat, 474, 616 
\bibitem[Nandra et al.(2013)]{nandra13} Nandra, K., Barret, D., Barcons, X., et al.\ 2013, arXiv:1306.2307 
\bibitem[Niida et al.(2016)]{niida16} Niida, M., Nagao, T., Ikeda, H., et al.\ 2016, \apj, 832, 208 
\bibitem[Oesch et al.(2016)]{oesch16} Oesch, P.~A., Brammer, G., van Dokkum, P.~G., et al.\ 2016, \apj, 819, 129 
\bibitem[Oesch et al.(2018)]{oesch18} Oesch, P.~A., Bouwens, R.~J., Illingworth, G.~D., Labb{\'e}, I., \& Stefanon, M.\ 2018, \apj, 855, 105 
\bibitem[Oke \& Gunn(1983)]{oke83} Oke, J.~B., \& Gunn, J.~E.\ 1983, \apj, 266, 713 
\bibitem[Ono et al.(2018)]{ono18} Ono, Y., Ouchi, M., Harikane, Y., et al.\ 2018, \pasj, 70, S10 
\bibitem[Onoue et al.(2017)]{onoue17} Onoue, M., Kashikawa, N., Willott, C.~J., et al.\ 2017, \apjl, 847, L15 
\bibitem[Parsa et al.(2018)]{parsa18} Parsa, S., Dunlop, J.~S., \& McLure, R.~J.\ 2018, \mnras, 474, 2904 
\bibitem[Planck Collaboration(2016)]{planck16} Planck Collaboration, Ade, P.~A.~R., Aghanim, N., et al.\ 2016, \aap, 594, A13 
\bibitem[Planck Collaboration et al.(2018)]{planck18} Planck Collaboration, Aghanim, N., Akrami, Y., et al.\ 2018, arXiv:1807.06209 
\bibitem[Ricci et al.(2017)]{ricci17} Ricci, F., Marchesi, S., Shankar, F., La Franca, F., \& Civano, F.\ 2017, \mnras, 465, 1915 
\bibitem[Richards et al.(2002)]{richards02} Richards, G.~T., Fan, X., Newberg, H.~J., et al.\ 2002, \aj, 123, 2945 
\bibitem[Robertson et al.(2015)]{robertson15} Robertson, B.~E., Ellis, R.~S., Furlanetto, S.~R., \& Dunlop, J.~S.\ 2015, \apjl, 802, L19 
\bibitem[Ross et al.(2012)]{ross12} Ross, N.~P., Myers, A.~D., Sheldon, E.~S., et al.\ 2012, \apjs, 199, 3 
\bibitem[Schlegel et al.(1998)]{schlegel98} Schlegel, D.~J., Finkbeiner, D.~P., \& Davis, M.\ 1998, \apj, 500, 525 
\bibitem[Serjeant et al.(2000)]{serjeant00} Serjeant, S., Oliver, S., Rowan-Robinson, M., et al.\ 2000, \mnras, 316, 768 
\bibitem[Shen et al.(2018)]{shen18} Shen, Y., Wu, J., Jiang, L., et al.\ 2018, arXiv:1809.05584 
\bibitem[Shull et al.(2012)]{shull12} Shull, J.~M., Harness, A., Trenti, M., \& Smith, B.~D.\ 2012, \apj, 747, 100 
\bibitem[Songaila(2004)]{songaila04} Songaila, A.\ 2004, \aj, 127, 2598 
\bibitem[Songaila \& Cowie(2010)]{songaila10} Songaila, A., \& Cowie, L.~L.\ 2010, \apj, 721, 1448 
\bibitem[Stanway et al.(2005)]{stanway05} Stanway, E.~R., McMahon, R.~G., \& Bunker, A.~J.\ 2005, \mnras, 359, 1184 
\bibitem[Totani et al.(2006)]{totani06} Totani, T., Kawai, N., Kosugi, G., et al.\ 2006, \pasj, 58, 485 
\bibitem[Vito et al.(2016)]{vito16} Vito, F., Gilli, R., Vignali, C., et al.\ 2016, \mnras, 463, 348 
\bibitem[Yang et al.(2016)]{yang16} Yang, J., Wang, F., Wu, X.-B., et al.\ 2016, \apj, 829, 33 
\bibitem[York et al.(2000)]{york00} York, D.~G., Adelman, J., Anderson, J.~E., Jr., et al.\ 2000, \aj, 120, 1579 
\bibitem[Weigel et al.(2015)]{weigel15} Weigel, A.~K., Schawinski, K., Treister, E., et al.\ 2015, \mnras, 448, 3167 
\bibitem[Willott et al.(2010)]{willott10} Willott, C.~J., Delorme, P., Reyl{\'e}, C., et al.\ 2010, \aj, 139, 906 
\bibitem[Willott et al.(2009)]{willott09} Willott, C.~J., Delorme, P., Reyl{\'e}, C., et al.\ 2009, \aj, 137, 3541 
\bibitem[Worseck \& Prochaska(2011)]{worseck11} Worseck, G., \& Prochaska, J.~X.\ 2011, \apj, 728, 23 

\end{thebibliography}
\end{document}